\documentclass[apjl,iop,twocolumn]{aastex61}
\usepackage{comment}
\usepackage{ifthen}
\usepackage{graphicx}
\usepackage[switch]{lineno}

\newcommand{\forloop}[5][1]%
{%
\setcounter{#2}{#3}%
\ifthenelse{#4}%
	{%
	#5%
	\addtocounter{#2}{#1}%
	\forloop[#1]{#2}{\value{#2}}{#4}{#5}%
	}%
	{%
	}%
}%

\newcommand{\ctbd}[1]{}

\newcommand{\lc}{light curve}

\newcommand{\Lc}{Light curve}

\newcommand{\magn}{magnitude}

\newcommand{\cmd}{color-magnitude diagram}

\newcommand{\band}[1]{\ensuremath{#1}~band}

\newcommand{\kms}{\ensuremath{\rm km\,s^{-1}}}
\newcommand{\ms}{\ensuremath{\rm m\,s^{-1}}}

\newcommand{\gcmc}{\ensuremath{\rm g\,cm^{-3}}}
\newcommand{\ergscmsq}{\ensuremath{\rm erg\,s^{-1}\,cm^{-2}}}

\newcommand{\logg}{\ensuremath{\log{g}}}
\newcommand{\vsini}{\ensuremath{v \sin{i}}}
\newcommand{\feh}{\ensuremath{\rm [Fe/H]}}

\newcommand{\rsun}{\ensuremath{R_\sun}}
\newcommand{\msun}{\ensuremath{M_\sun}}
\newcommand{\lsun}{\ensuremath{L_\sun}}

\newcommand{\rstar}{\ensuremath{R_\star}}
\newcommand{\mstar}{\ensuremath{M_\star}}
\newcommand{\lstar}{\ensuremath{L_\star}}

\newcommand{\teffstar}{\ensuremath{T_{\rm eff\star}}}
\newcommand{\rhostar}{\ensuremath{\rho_\star}}
\newcommand{\loggstar}{\ensuremath{\log{g_{\star}}}}

\newcommand{\rpl}{\ensuremath{R_{p}}}
\newcommand{\mpl}{\ensuremath{M_{p}}}

\newcommand{\rhopl}{\ensuremath{\rho_{p}}}

\newcommand{\arstar}{\ensuremath{a/\rstar}}
\newcommand{\zrstar}{\ensuremath{\zeta/\rstar}}

\newcommand{\rjup}{\ensuremath{R_{\rm J}}}
\newcommand{\mjup}{\ensuremath{M_{\rm J}}}

\newcommand{\refsecl}[1]{\mbox{Section \ref{sec:#1}}}

\newcommand{\reftabl}[1]{Table~\ref{tab:#1}}

\newcommand{\loopand}{\ifnum\value{planetcounter}=2 and \else\fi}
\newcommand{\loopcomma}{\ifnum\value{planetcounter}<2 ,\else. \fi}
\newcommand{\loopcommanoperiod}{\ifnum\value{planetcounter}<2 ,\else \space\fi}
\newcommand{\loopcommanospace}{\ifnum\value{planetcounter}<2 ,\else \fi}

\newcommand{\hatcurISOmisochronecirc}{\ensuremath{0.6785_{-0.0079}^{+0.0299}}} 
\newcommand{\hatcurISOrisochronecirc}{\ensuremath{0.6701_{-0.0032}^{+0.0041}}} 
\newcommand{\hatcurISOloggisochronecirc}{\ensuremath{4.615\pm0.013}}   
\newcommand{\hatcurISOteffisochronecirc}{\ensuremath{4508\pm43}}       
\newcommand{\hatcurISOzfehisochronecirc}{\ensuremath{-0.059\pm0.036}}  
\newcommand{\hatcurISOageisochronecirc}{\ensuremath{11.1_{-6.9}^{+1.1}}} 
\newcommand{\hatcurXdistredisochronecirc}{\ensuremath{202.93\pm0.97}}  
\newcommand{\hatcurRVeccenisochroneeccen}{\ensuremath{0.013\pm0.013}}   
\newcommand{\hatcurRVeccentwosiglimisochroneeccen}{\ensuremath{<0.041}} 
\newcommand{\hatcurSMEiteffempiricalcirc}{\ensuremath{4514\pm50}}      
\newcommand{\hatcurSMEizfehempiricalcirc}{\ensuremath{-0.140\pm0.080}} 
\newcommand{\hatcurSMEizfehshortempiricalcirc}{\ensuremath{-0.14}}     
\newcommand{\hatcurSMEiloggempiricalcirc}{\ensuremath{4.67\pm0.10}}    
\newcommand{\hatcurSMEivsinempiricalcirc}{\ensuremath{0.0\pm2.0}}      
\newcommand{\hatcurSMEivmacempiricalcirc}{\ensuremath{nff\pmnff}}      
\newcommand{\hatcurSMEivmicempiricalcirc}{\ensuremath{nff\pmnff}}      
\newcommand{\hatcurISOmempiricalcirc}{\ensuremath{0.673_{-0.014}^{+0.020}}} 
\newcommand{\hatcurISOmnoisorestrictempiricalcirc}{\ensuremath{0.614\pm0.055}} 
\newcommand{\hatcurISOrempiricalcirc}{\ensuremath{0.6726\pm0.0069}}    
\newcommand{\hatcurISOrnoisorestrictempiricalcirc}{\ensuremath{0.6720\pm0.0075}} 
\newcommand{\hatcurPPmlongempiricalcirc}{\ensuremath{0.711\pm0.038}}   
\newcommand{\hatcurPPrlongempiricalcirc}{\ensuremath{1.072\pm0.015}}   
\newcommand{\hatcurPPteffempiricalcirc}{\ensuremath{1015.0_{-5.7}^{+14.9}}} 
\newcommand{\hatcurXdistrednoisorestrictempiricalcirc}{\ensuremath{203.46\pm1.00}} 
\newcommand{\hatcurCCra}{\ensuremath{07^{\mathrm h}53^{\mathrm m}55.9828{\mathrm s}}}                   
\newcommand{\hatcurCCdec}{\ensuremath{23{\arcdeg}56{\arcmin}17.6117{\arcsec}}}                  
\newcommand{\hatcurCCtwomass}{07535598+2356176}     
\newcommand{\hatcurCCgsc}{1925-01046}                 
\newcommand{\hatcurCCgaiadrtwo}{675443053940533760} 
\newcommand{\hatcurCCtassmv}{\ensuremath{13.937\pm0.030}} 
\newcommand{\hatcurCCtassmB}{\ensuremath{15.148\pm0.030}} 
\newcommand{\hatcurCCtassmg}{\ensuremath{14.578\pm0.020}} 
\newcommand{\hatcurCCtassmr}{\ensuremath{13.421\pm0.050}} 
\newcommand{\hatcurCCtassmi}{\ensuremath{13.52\pm0.82}}   
\newcommand{\hatcurCCparallax}{\ensuremath{4.918\pm0.023}} 
\newcommand{\hatcurCCgaiamG}{\ensuremath{13.54420\pm0.00070}} 
\newcommand{\hatcurCCgaiamBP}{\ensuremath{14.2484\pm0.0028}} 
\newcommand{\hatcurCCgaiamRP}{\ensuremath{12.7429\pm0.0017}} 
\newcommand{\hatcurCCtwomassJmag}{\ensuremath{11.750\pm0.022}} 
\newcommand{\hatcurCCtwomassHmag}{\ensuremath{11.210\pm0.023}} 
\newcommand{\hatcurCCtwomassKmag}{\ensuremath{11.019\pm0.018}} 
\newcommand{\hatcurCCWonemag}{\ensuremath{10.956\pm0.025}} 
\newcommand{\hatcurCCWtwomag}{\ensuremath{11.021\pm0.021}} 
\newcommand{\hatcurCCWthreemag}{\ensuremath{10.86\pm0.10}} 
\newcommand{\hatcurLCrprstar}{\ensuremath{0.1644\pm0.0015}} 
\newcommand{\hatcurLCbsq}{\ensuremath{0.045_{-0.026}^{+0.021}}} 
\newcommand{\hatcurLCimp}{\ensuremath{0.212_{-0.076}^{+0.045}}} 
\newcommand{\hatcurLCzeta}{\ensuremath{26.95\pm0.21}}     
\newcommand{\hatcurLCdur}{\ensuremath{0.08695\pm0.00053}} 
\newcommand{\hatcurLCingdur}{\ensuremath{0.01279\pm0.00033}} 
\newcommand{\hatcurLCP}{\ensuremath{2.29840551\pm0.00000052}} 
\newcommand{\hatcurLCPshort}{\ensuremath{2.2984}}         
\newcommand{\hatcurLCT}{\ensuremath{2456614.20355\pm0.00014}} 
\newcommand{\hatcurLBig}{\ensuremath{0.718\pm0.097}}      
\newcommand{\hatcurLBiig}{\ensuremath{0.25\pm0.11}}       
\newcommand{\hatcurLBir}{\ensuremath{0.49\pm0.14}}        
\newcommand{\hatcurLBiir}{\ensuremath{0.41\pm0.15}}       
\newcommand{\hatcurLBii}{\ensuremath{0.337\pm0.073}}      
\newcommand{\hatcurLBiii}{\ensuremath{0.34\pm0.14}}       
\newcommand{\hatcurISOm}{\ensuremath{0.6785_{-0.0079}^{+0.0299}}} 
\newcommand{\hatcurISOmlong}{\ensuremath{0.6785_{-0.0079}^{+0.0299}}} 
\newcommand{\hatcurISOr}{\ensuremath{0.6701_{-0.0032}^{+0.0041}}} 
\newcommand{\hatcurISOrlong}{\ensuremath{0.6701_{-0.0032}^{+0.0041}}} 
\newcommand{\hatcurISOrho}{\ensuremath{3.165_{-0.046}^{+0.187}}} 
\newcommand{\hatcurISOlogg}{\ensuremath{4.615\pm0.013}}   
\newcommand{\hatcurISOlum}{\ensuremath{0.1663\pm0.0062}}  
\newcommand{\hatcurISOteff}{\ensuremath{4508\pm43}}       
\newcommand{\hatcurISOzfeh}{\ensuremath{-0.059\pm0.036}}  
\newcommand{\hatcurISOage}{\ensuremath{11.1_{-6.9}^{+1.1}}} 
\newcommand{\hatcurISOspec}{K}                            
\newcommand{\hatcurRVK}{\ensuremath{143.4\pm8.4}}         
\newcommand{\hatcurRVjitterA}{\ensuremath{14.3\pm7.9}}    
\newcommand{\hatcurRVjitterB}{\ensuremath{0\pm25}}        
\newcommand{\hatcurRVjitterC}{\ensuremath{0\pm59}}        
\newcommand{\hatcurPPi}{\ensuremath{88.73_{-0.27}^{+0.47}}} 
\newcommand{\hatcurPPlogg}{\ensuremath{3.192\pm0.029}}    
\newcommand{\hatcurPPar}{\ensuremath{9.600_{-0.047}^{+0.185}}} 
\newcommand{\hatcurPParel}{\ensuremath{0.02996_{-0.00012}^{+0.00043}}} 
\newcommand{\hatcurPPrho}{\ensuremath{0.727\pm0.051}}     
\newcommand{\hatcurPPmlong}{\ensuremath{0.724\pm0.043}}   
\newcommand{\hatcurPPrlong}{\ensuremath{1.072\pm0.012}}   
\newcommand{\hatcurPPmrcorr}{\ensuremath{-0.04}}          
\newcommand{\hatcurPPteff}{\ensuremath{1027.8\pm8.2}}     
\newcommand{\hatcurPPtheta}{\ensuremath{0.0590\pm0.0035}} 
\newcommand{\hatcurPPfluxavg}{\ensuremath{2.515\pm0.080}} 
\newcommand{\hatcurPPfluxavgdim}{\ensuremath{8}}          
\newcommand{\hatcurXAv}{\ensuremath{0.180\pm0.075}}       
\newcommand{\hatcurXdistred}{\ensuremath{202.93\pm0.97}}  
\newcommand{\hatcurCCpmra}{\ensuremath{-23.655\pm0.039}}  
\newcommand{\hatcurCCpmdec}{\ensuremath{-1.467\pm0.022}}  

\newcommand{\hatcur}{HAT-P-68}
\newcommand{\hatcurb}{HAT-P-68b}

\newcommand{\hatcurSMEversion}{i}                                       
\newcommand{\hatcurSMEteff}{\ifthenelse{\equal{\hatcurSMEversion}{i}}{\hatcurSMEiteffempiricalcirc}{\hatcurSMEiiteffempiricalcirc}}
\newcommand{\hatcurSMEzfeh}{\ifthenelse{\equal{\hatcurSMEversion}{i}}{\hatcurSMEizfehempiricalcirc}{\hatcurSMEiizfehempiricalcirc}}
\newcommand{\hatcurSMEzfehshort}{\ifthenelse{\equal{\hatcurSMEversion}{i}}{\hatcurSMEizfehshortempiricalcirc}{\hatcurSMEiizfehshortempiricalcirc}}
\newcommand{\hatcurSMElogg}{\ifthenelse{\equal{\hatcurSMEversion}{i}}{\hatcurSMEiloggempiricalcirc}{\hatcurSMEiiloggempiricalcirc}}
\newcommand{\hatcurSMEvsin}{\ifthenelse{\equal{\hatcurSMEversion}{i}}{\hatcurSMEivsinempiricalcirc}{\hatcurSMEiivsinempiricalcirc}}
\newcommand{\hatcurSMEvmac}{\ifthenelse{\equal{\hatcurSMEversion}{i}}{\hatcurSMEivmacempiricalcirc}{\hatcurSMEiivmacempiricalcirc}}
\newcommand{\hatcurSMEvmic}{\ifthenelse{\equal{\hatcurSMEversion}{i}}{\hatcurSMEivmicempiricalcirc}{\hatcurSMEiivmicempiricalcirc}}

\newcounter{planetcounter}

\newboolean{emulateapj}
\setboolean{emulateapj}{true}

\newboolean{rvtablelong}
\setboolean{rvtablelong}{true}

\newboolean{astroph}
\setboolean{astroph}{true}

\shortauthors{Lindor et al.}
\shorttitle{\hatcur\lowercase{b}}
\ifthenelse{\boolean{emulateapj}}{
    
}{
    
}

\begin{document}

\title{
\hatcur\lowercase{b}: A Transiting Hot Jupiter Around a K5 Dwarf Star \footnote{
Based on observations obtained with the Hungarian-made Automated
Telescope Network. Based in part on observations made with the Keck-I
telescope at Mauna Kea Observatory, HI (Keck time awarded through NASA
programs N133Hr and N169Hr). Based in part on observations obtained with
the Tillinghast Reflector 1.5\,m telescope and the 1.2\,m telescope,
both operated by the Smithsonian Astrophysical Observatory at the Fred
Lawrence Whipple Observatory in Arizona. Based on radial velocities obtained 
with the Sophie spectrograph mounted on the 1.93\,m telescope at Observatoire de Haute-Provence.
}}

\correspondingauthor{Bethlee M. Lindor}
\email{blindor@uw.edu}

\author{Bethlee M. Lindor}
\affil{Astronomy Department, University of Washington, Seattle, WA 98195, USA}
\affil{NSF Graduate Student Research Fellow}
\affil{Department of Astrophysical Sciences, Princeton University, NJ 08544, USA}

\author[0000-0001-8732-6166]{Joel D. Hartman}
\affil{Department of Astrophysical Sciences, Princeton University, NJ 08544, USA}

\author[0000-0001-7204-6727]{G\'asp\'ar \'A. Bakos}
\altaffiliation{Packard Fellow}
\affil{Department of Astrophysical Sciences, Princeton University, NJ 08544, USA}
\affil{MTA Distinguished Guest Fellow, Konkoly Observatory, Hungary}

\author[0000-0002-0628-0088]{Waqas Bhatti}
\affil{Department of Astrophysical Sciences, Princeton University, NJ 08544, USA}

\author{Zoltan Csubry}
\affil{Department of Astrophysical Sciences, Princeton University, NJ 08544, USA}

\author[0000-0003-4464-1371]{Kaloyan Penev}
\affil{Department of Physics, University of Texas at Dallas, Richardson, TX 75080, USA}

\author[0000-0001-6637-5401]{Allyson Bieryla}
\affiliation{Harvard-Smithsonian Center for Astrophysics, 60 Garden St, Cambridge, MA 02138, USA}

\author[0000-0001-9911-7388]{David~W.~Latham}
\affiliation{Harvard-Smithsonian Center for Astrophysics, 60 Garden St, Cambridge, MA 02138, USA}

\author[0000-0002-5286-0251]{Guillermo Torres}
\affiliation{Harvard-Smithsonian Center for Astrophysics, 60 Garden St, Cambridge, MA 02138, USA}

\author[0000-0003-1605-5666]{Lars~A.~Buchhave}
\affiliation{DTU Space, National Space Institute, Technical University of Denmark, Elektrovej 328, DK-2800 Kgs. Lyngby, Denmark}

\author{G\'eza Kov\'acs}
\affiliation{Konkoly Observatory of the Hungarian Academy of Sciences, Budapest, Hungary}

\author[0000-0002-0455-9384]{Miguel de Val-Borro}
\affil{Astrochemistry Laboratory, Goddard Space Flight Center, NASA, 8800 Greenbelt Rd, Greenbelt, MD 20771, USA}

\author[0000-0001-8638-0320]{Andrew~W.~Howard}
\affil{Department of Astronomy, California Institute of Technology, Pasadena, CA, USA}

\author[0000-0002-0531-1073]{Howard~Isaacson}
\affil{Department of Astronomy, University of California, Berkeley, CA, USA}

\author[0000-0003-3504-5316]{Benjamin~J.~Fulton}
\affil{Department of Astronomy, California Institute of Technology, Pasadena, CA, USA}
\affil{IPAC-NASA Exoplanet Science Institute, Pasadena, CA, USA}

\author[0000-0001-8388-8399]{Isabelle Boisse}
\affil{Aix Marseille Universit\'e, CNRS, CNES, LAM (Laboratoire d'Astrophysique de Marseille), Marseille, France}

\author[0000-0002-3586-1316]{Alexandre Santerne}
\affil{Aix Marseille Universit\'e, CNRS, CNES, LAM (Laboratoire d'Astrophysique de Marseille), Marseille, France}

\author{Guillaume H\'ebrard}
\affil{Institut d’Astrophysique de Paris, UMR7095 CNRS, Universit\'e Pierre \& Marie Curie, 98bis boulevard Arago, 75014 Paris, France}

\author[0000-0002-0697-6050]{T\'am\'as Kov\'acs}
\affiliation{Institute of Physics, Eötvös University, 1117 Budapest, Hungary}

\author[0000-0003-0918-7484]{Chelsea~X.~Huang}
\affiliation{Department of Physics, and Kavli Institute for Astrophysics and Space Research, Massachusetts Institute of Technology, Cambridge, MA 02139, USA}

\author{Jack~Dembicky}
\affiliation{Apache Point Observatory, Sunspot, NM 88349, USA}

\author[0000-0002-7061-6519]{Emilio Falco}
\affiliation{Harvard-Smithsonian Center for Astrophysics, 60 Garden St, Cambridge, MA 02138, USA}

\author[0000-0002-0885-7215]{Mark E.~Everett}
\affiliation{National Optical-Infrared Astronomy Research Laboratory, Tucson, AZ 85719 USA.}

\author[0000-0003-2159-1463]{Elliott~P.~Horch}
\altaffiliation{Adjunct Astronomer, Lowell Observatory}
\affiliation{Department of Physics, Southern Connecticut State University, 501 Crescent Street, New Haven, CT 06515, USA}

\author{J\'ozsef L\'az\'ar}
\affil{Hungarian Astronomical Association, 1451 Budapest, Hungary}

\author{Istv\'an Papp}
\affil{Hungarian Astronomical Association, 1451 Budapest, Hungary}

\author{P\'al S\'ari}
\affil{Hungarian Astronomical Association, 1451 Budapest, Hungary}


\begin{abstract}

\setcounter{footnote}{1}
We report the discovery by the ground-based HATNet survey of the transiting exoplanet 
\hatcurb{}, which has a mass of \hatcurPPmlong\,\mjup, and radius of \hatcurPPrlong\,\rjup. 
The planet is in a circular $P=\hatcurLCPshort$\,-day orbit around
a moderately bright V = \hatcurCCtassmv\,\magn\ \hatcurISOspec\ dwarf star of mass 
\hatcurISOmempiricalcirc\,\msun, and radius \hatcurISOrempiricalcirc\,\rsun. 
The planetary nature of this system is confirmed through follow-up transit photometry obtained with the FLWO~1.2\,m telescope, high-precision RVs measured using Keck-I/HIRES, FLWO~1.5\,m/TRES, and OHP~1.9\,m/Sophie, and high-spatial-resolution speckle imaging from WIYN~3.5\,m/DSSI.

\hatcur\ is at an ecliptic latitude of $+3^{\circ}$ and outside the field of view of both the NASA {\em TESS} primary mission and the {\em K2} mission.
The large transit depth of $0.036$\,mag ($r$-band) makes \hatcurb{} a 
promising
target for atmospheric characterization via transmission spectroscopy.

\setcounter{footnote}{0}
\end{abstract}

\keywords{
    planetary systems ---
    stars: individual (
\hatcur{},
GSC~\hatcurCCgsc{}
) 
    techniques: spectroscopic, photometric
}


\section{Introduction}
\label{sec:introduction}


The first detection of a planet orbiting a star besides our own \citep{
mayor:1995:exo} sparked a new era of astronomy and planetary science, 
and made
the discovery and characterization of extra-solar planets a focal point of 
observational research in astrophysics. Among the various methods available,
transit photometry 
has produced the largest yield of exoplanets to date, and has also proven to be the most sensitive method for discovering small planets\footnote{NASA Exoplanet Archive accessed Sept.\ 2020; \url{http://exoplanetarchive.ipac.caltech.edu}}.
Additionally, transiting exoplanets (TEPs) offer the unique opportunity to study the 
physical properties of planets outside the Solar System, and how these properties 
depend on those of their parent stars. 
Combining transit time-series data 
with measurements of the radial velocity (RV) orbital wobble of the host star 
provides the masses and radii of planetary objects -- that is, if 
the stellar mass and radius can be determined through other means.
Furthermore, follow-up observations of these systems allow us to study the structure 
and composition of the planetary atmospheres through transmission 
spectroscopy \cite[e.g.][]{char:2002:atm}, and to measure the orbital 
eccentricity and obliquity \citep[e.g.][]{morton:2014}. 
These capabilities make TEPs one of the most reliable sources for testing current 
models of planetary formation and evolution.

Many wide-field ground-based surveys have 
been productive
in detecting TEPs, 
with the largest yields coming from WASP \citep{pollacco:2006}, HATSouth \citep{bakos:2013:hatsouth} and
HATNet \citep{bakos:2004:hatnet}.
The sample of exoplanets discovered by these surveys
is highly biased towards giant planets at 
short orbital distances to their host stars \citep[e.g.][]{gaudi:2005}. 
These hot Jupiters (HJs) initially shattered our understanding of planetary formation. 
Space surveys like the all-sky Transiting Exoplanet Survey Satellite 
\citep[TESS;][]{ricker:2014} -- joining the legacy of \textit{Kepler} \citep{borucki:2010}, 
\textit{K2} \citep{howell:2014}, and \textit{CoRot} \citep{auvergne:2009} -- are better equipped 
to identify objects with a wider range of sizes at a wide range of orbital distances to their host stars \citep[e.g.,][]{nielsen:2020}.

Yet, discoveries of planets with orbital periods shorter than 10 days provide advantages to 
resolving current theoretical challenges in the field \citep[see][]{dawson:2018}. 
For instance, explaining the inflated radii of HJs remains a theoretical puzzle \citep[e.g.][and references therein]{sestovic:2018} that may be elucidated by building up a larger sample objects to disentangle the effects of age, orbital separation, irradiation, composition and mass on the radii of these planets \citep[e.g.,][]{hartman:2016:hat65hat66}. Explaining the origin of these planets 
as well as understanding how they evolve via planet-star interactions are subjects 
that can be better addressed with a larger sample of objects. 

The HATNet survey searches for planets transiting moderately bright stars by 
utilizing six small telephoto lenses on robotic mounts. 
Specifically, HATNet has two stations with multiple 11\,cm telescopes; one of which 
is located at the Smithsonian Astrophysical Observatory's Fred Lawrence Whipple 
Observatory (FLWO) in Arizona, while the other is atop the Mauna Kea Observatory 
(MKO) in Hawaii. \citet{bakos:2018:book} provides a recent review of the HATNet and HATSouth projects.

Here we present the discovery by the HATNet survey of a transiting, short-period, gas-giant planet 
around a \hatcurISOspec\ dwarf star. Section~\ref{sec:obs} summarizes the 
observational data that led to the discovery, as well as various follow-up studies 
performed for \hatcur. This involved photometric and spectroscopic observations, 
and high resolution imaging. In Section~\ref{sec:analysis}, we analyze the data
to rule out false positive scenarios and determine the 
best-fit stellar and planetary parameters. 
We discuss our results in Section~\ref{sec:discussion}.

\section{Observations}
\label{sec:obs}

We have used a number of observations to aid our understanding of \hatcur{}, 
and to confirm the existence of an extra-solar planet in the system. 
These observations include discovery light curves obtained by the HATNet survey, 
ground-based follow-up transit light curves, high-resolution spectra and associated RVs, high-spatial resolution imaging, and 
catalog broad-band photometry and astrometry. 
We describe the observations collected by our team in the following sections. 
See Tables~\ref{tab:specobs} and \ref{tab:photobs} for brief summaries of all 
the spectroscopic and photometric observations collected for \hatcur{}.

\subsection{Photometric Detection}
\label{sec:detection}

\begin{figure}[t]
\epsscale{1.25}
\plotone{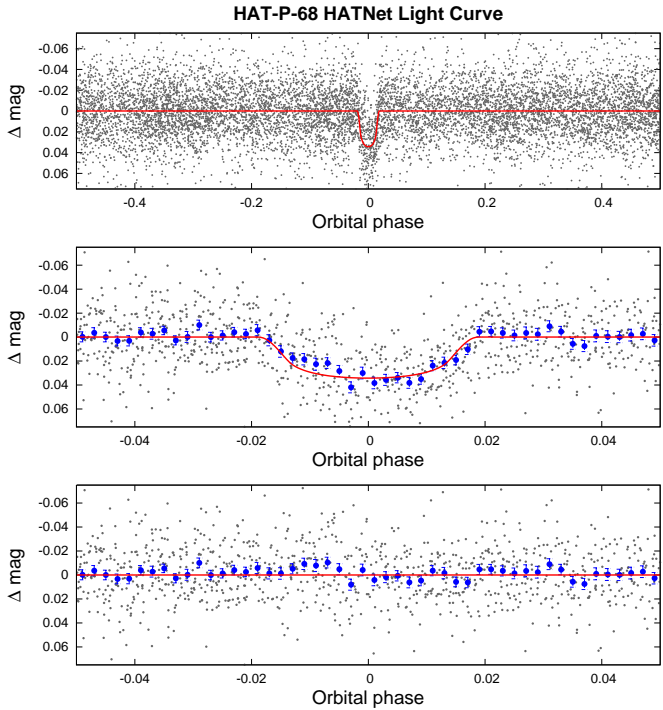}
\caption{
        Discovery HATNet transit \lc{} phase-folded with a period of \hatcurLCPshort\,days. {\em Top:} The full unbinned instrumental \band{r} \lc{}. The gray points show the individual measurements, while the solid red line shows the best-fit transit model. {\em Middle:} Same as the top panel, here we restrict the horizontal range of the plot to better display the transit. The filled blue circles show the \lc{} averaged in phase using a bin-size of 0.002. {\em Bottom:} The residuals from the best-fit transit model.\\
\label{fig:hatnet}}
\end{figure}

\begin{figure}[t]
\epsscale{1.25}
\plotone{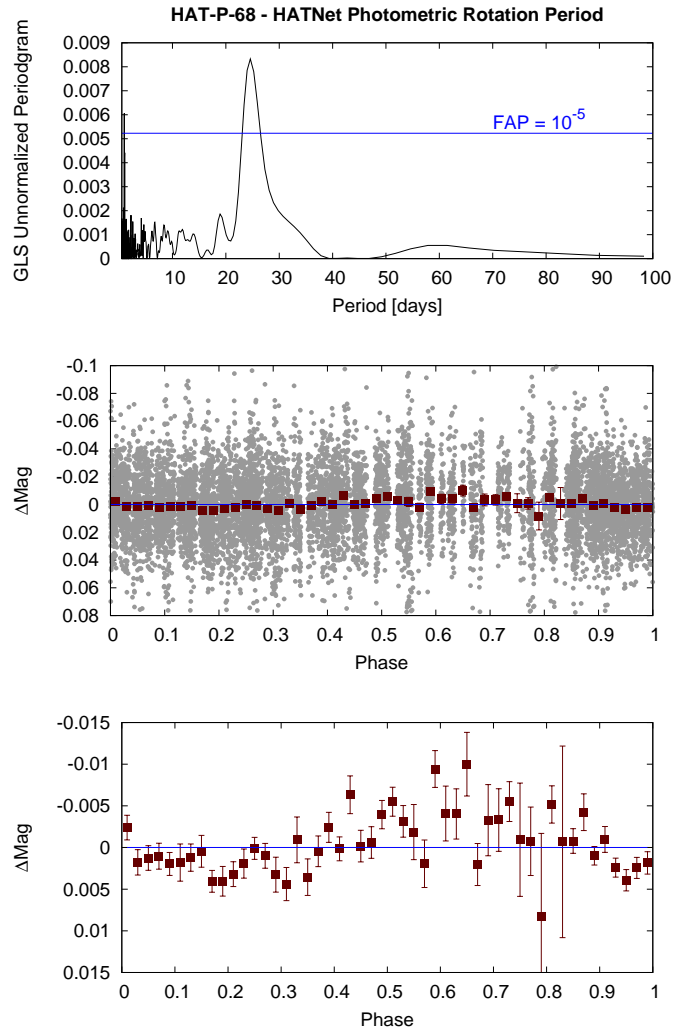}
\caption{
        Detection of a $P = 24.593 \pm 0.064$\,days photometric rotation period signal in the HATNet \lc{} of \hatcur. {\em Top:} The Generalized Lomb-Scargle (GLS) periodogram of the HATNet \lc{} after subtracting the best-fit transit model. The horizontal blue line shows a bootstrap-calibrated $10^{-5}$ false alarm probability (FAP) level. {\em Middle:} The HATNet \lc{} phase-folded at the peak GLS period. The gray points show the individual photometric measurements, while the dark red filled squares show the observations binned in phase with a bin size of 0.02. {\em Bottom:} Same as the middle panel, here we restrict the vertical range of the plot to better show the variation seen in the phase-binned measurements. We only show the phase binned measurements in this case as the variation in the un-binned points exceeds the vertical axis range of the plot.\\
\label{fig:gls}}
\end{figure}

Observations of a field containing \hatcur{} were carried out between 
2011 November and 2012 May 
by 
the HAT-5 and HAT-8 instruments 
located at FLWO and MKO, respectively. 
A total of 5867 and 3034 exposures of 3 minutes were obtained with 
each device through a Sloan $r$ filter, after which the images were 
reduced to trend-filtered light curves following \citet{bakos:2010:hat11}. 
Here we 
used
the Trend-Filtering Algorithm \citep[TFA;][]{kovacs:2005:TFA} in 
signal-detection mode 
(i.e., the filtering was done before identifying the transit signal, and no attempt was made to preserve the shape of the transit during the filtering process).
The final point-to-point precision for the HATNet
\lc{} of \hatcur{} is 2.4\%.  

We searched the light curves from the aforementioned field for periodic 
box-shaped transit events using the Box Least Squares method 
\citep[BLS;][]{kovacs:2002:BLS}, and detected 3.6\% deep transits with a 
period of \hatcurLCPshort\,days in the \lc{} of \hatcur{}. This 
detection 
prompted additional photometric and spectroscopic follow-up observations, as described
in the subsections 
below. Figure~\ref{fig:hatnet} shows the HATNet \lc{} phase folded at the 
period identified with BLS, together with our best-fit transit model. 
The differential photometry data are made available in Table~\ref{tab:phfu}.

After subtracting the best-fit primary transit model from the HATNet light curve, 
we used BLS to search the residuals for additional periodic transit signals. 
No other significant transit signals were identified. We can 
place an approximate upper limit of 1\% on the depth of any other periodic transit 
signals in the \lc{} with periods 
shorter 
than $\sim10$\,days.

To supplement 
the search
for periodic transit signals, we also searched the HATNet 
\lc{} residuals for sinusoidal periodic variations using the Generalized 
Lomb-Scargle (GLS) periodogram \citep{zechmeister:2009}. 
This search detected a $P = 24.593 \pm 0.064$\,day periodic quasi-sinusoidal signal, from which we computed a bootstrap-calibrated false alarm probability of $10^{-10.3}$ and a periodogram signal-to-noise ratio of $36$ as described in \citet{hartman:2016:vartools}. The GLS periodogram and phase-folded \lc{} are shown in Figure~\ref{fig:gls}. 
We provisionally identify this as the photometric rotation period of the star, 
and note that the period and amplitude are in line with other mid K dwarf main sequence stars \citep[e.g.,][]{hartman:2011:kmdwarf}.

\subsection{Reconnaissance Spectroscopy}
\label{sec:obsspec}


Initial reconnaissance spectroscopy observations of \hatcur{} were
obtained using the Astrophysical Research Consortium Echelle
Spectrometer \citep[ARCES;][]{wang:2003} on the ARC~3.5\,m telescope
located at Apache Point Observatory (APO) in New Mexico. Using this facility,
we obtained three $\Delta \lambda$/$\lambda \equiv R = 18,000$
resolution spectra of \hatcur{} on UT 2012 Oct 30, 2012 Nov 7, and
2013 Mar 3. These had exposure times of 3600\,s, 2740\,s, and 2740\,s,
respectively, yielding signal-to-noise ratios per resolution element
near 5180\,\AA\ of 32.3, 25.6, and 26.8, respectively. The \'echelle
images were reduced to wavelength-calibrated spectra following
\citet{hartman:2015:hat50hat53}. 

We applied the Stellar Parameter Classification \citep[SPC;][]{buchhave:2012:spc} 
method on the \'echelle images to measure the RV and 
atmospheric parameters for the stellar host. In particular, this pipeline derives
the effective temperature (\teffstar), surface gravity (\logg), metallicity (\feh)
and projected equatorial rotation velocity (\vsini). 
Based on the three ARCES observations we
estimated $\teffstar = 4500 \pm 50$\,K, $\logg = 4.62 \pm 0.10$ (cgs),
$\feh = -0.14 \pm 0.08$ and $\vsini = 2.5 \pm 0.5$\,\kms. 

We caution that the uncertainties based on this analysis are likely underestimated compared to the values reported in Section~\ref{sec:starparams} based on an SPC analysis of Keck-I/HIRES observations. 
The three RV measurements 
were consistent with no variation, with a mean value of $-8.69$\,\kms, and a standard deviation of $0.43$\,\kms, comparable to the systematic uncertainties in the wavelength calibration. 
We note that the cross-correlation functions were consistent with a single K dwarf star, 
with no evidence of a second set of absorption lines present in the spectra.

\subsection{High RV-Precision Spectroscopy}
\label{sec:highspec}

\begin{figure}[t]
\epsscale{1.25}
\plotone{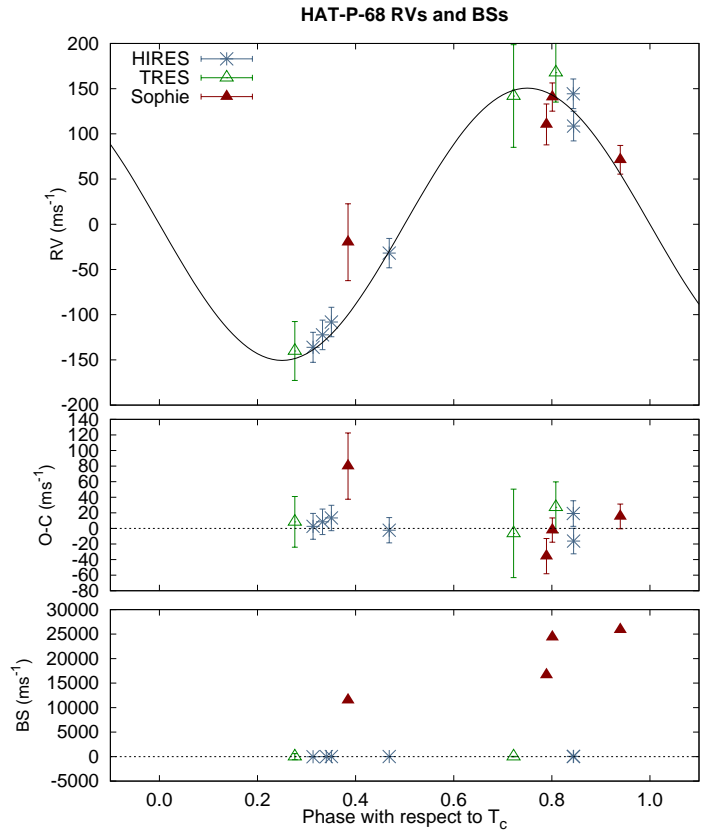}
\caption{
    \emph{Top:} High-precision RV measurements from 
    FLWO~1.5\,m/TRES, OHP~1.9\,m/Sophie, and Keck-I~10\,m/HIRES, 
    together with our best-fit orbit model, plotted as a function of orbital phase. 
    Phase zero corresponds to the time of mid transit. The center-of-mass
    velocity has been subtracted. The error bars include the jitter which is varied 
    independently for each instrument in the fit. 
    \emph{Middle:} RV $O\!-\!C$ residuals from the best-fit model, plotted as a function of phase. 
    \emph{Bottom:} Spectral line bisector spans (BSs) plotted as a function of phase. 
    Note the different vertical scales of the three panels.\\
\label{fig:rvbis}}
\end{figure}


Following the reconnaissance, we obtained higher resolution, and higher RV-precision
spectroscopic observations of \hatcur{} to further characterize it. 
To carry out these observations we used the Tillinghast Reflector Echelle Spectrograph
\citep[TRES;][]{furesz:2008} on the 1.5\,m Tillinghast Reflector at
FLWO, the Sophie spectrograph \citep{bouchy:2009} on the Observatoire
de Haute-Provence (OHP)~1.93\,m in France, and HIRES \citep{vogt:1994}
on the Keck-I~10\,m at MKO together with its I$_2$ absorption cell. 
The measured RVs and spectral line bisector spans (BSs) from these three
facilities are provided in Table~\ref{tab:rvs} and plotted in Figure~\ref{fig:rvbis}.

A total of 3 TRES spectra were obtained on UT 2012 Nov 23, 2013 Mar 1,
and 2013 Oct 11 at a resolution of $R = 44,000$ and were reduced to
high precision RVs and BSs following
\citet{bieryla:2014:hat49}, and to atmospheric stellar parameters
using SPC. 

A total of four $R = 39,000$ spectra were obtained with
Sophie on UT 2013 Oct 31, 2013 Nov 1, and 2013 Nov 6, and were reduced
to high-precision RVs and BSs following \citet{boisse:2013:hat4243}.  

A total of six $R = 55,000$ spectra were obtained through an I$_2$ 
cell with HIRES on UT 2013 Oct 19, 2013 Dec 11--12, and 2015 Nov 26--28. 
An I$_2$-free template observation was obtained on UT 2013 Oct 19. 
These data were collected and reduced following standard procedures of the 
California Planet Search \citep[CPS;][]{howard:2010:cps}, 
including computation of RVs using a method descended from 
\citet{butler:1996}, and BSs following \citet{torres:2007:hat3}. 
We also applied SPC to the I$_{2}$-free template to obtain high precision 
atmospheric parameters of the host star.
Note that while we did not have an RV observation for this observation, 
we did compute a BS measurement from the blue orders of the echelle for it.

As seen in Figure~\ref{fig:rvbis}, the RVs from TRES, Sophie and HIRES 
exhibited a clear Keplerian
orbital variation in phase with the ephemeris from the photometric
transits. We also find that the BSs from HIRES show almost no variation.
The TRES BS values had several hundred \ms\ uncertainties, 
and the Sophie values varied by many \kms, in both cases due to 
significant sky contamination that affected the shapes of the cross correlation functions (CCFs).

\ifthenelse{\boolean{emulateapj}}{
    \begin{deluxetable*}{llrcccr}
}{
    \begin{deluxetable}{llrcccr}
}
\tablewidth{0pc}
\tabletypesize{\scriptsize}
\tablecaption{
    Summary of Spectroscopic Observations
    \label{tab:specobs}
}
\tablehead{
    \multicolumn{1}{c}{Telescope/Instrument} &
    \multicolumn{1}{c}{UT Date(s)} &
    \multicolumn{1}{c}{\# Spectra} &
    \multicolumn{1}{c}{Resolution} &
    \multicolumn{1}{c}{S/N Range\tablenotemark{a}} &
    \multicolumn{1}{c}{$\gamma_{RV}$\tablenotemark{b}} &
    \multicolumn{1}{c}{RV Precision\tablenotemark{c}} \\
    \multicolumn{1}{c}{} &
    \multicolumn{1}{c}{} &
    \multicolumn{1}{c}{} &
    \multicolumn{1}{c}{$\Delta \lambda$/$\lambda$} &
    \multicolumn{1}{c}{} &
    \multicolumn{1}{c}{(\ms)} &
    \multicolumn{1}{c}{(\ms)} 
}
\startdata
~~~~APO~3.5\,m/ARCES & 2012 Oct--2013 Mar & 3 & 18000 & 26.8--32.3 & $-8600$ & 430 \\
~~~~FLWO~1.5\,m/TRES & 2012 Nov--2013 Nov & 3 & 44000 & 10--17 & $-7960$ & $16.8$ \\
~~~~OHP~1.9\,m/Sophie & 2013 Oct--Nov & 4 & 39000 & 26--43 & $-8890$ & $48.6$ \\
~~~~Keck-I~10\,m/HIRES & 2013 Oct--2015 Nov & 6 & 55000 & 81--115 & $\cdots$ & $12.5$ \\
\enddata
\tablenotetext{a}{
S/N per resolution element near 5180 \AA. This was not measured for all of the instruments.
}
\tablenotetext{b}{
For Sophie RV observations  
this is the zero-point RV from the best-fit orbit. For ARCES and TRES it is the mean value of the low-precision reconnaissance RV. Higher-precision RVs were measured from the TRES observations and used in the orbit fitting as well, however these are relative RV measurements that were not adjusted to an absolute standard.
}
\tablenotetext{c}{
For high-precision RV observations included in the orbit determination, this is the scatter in the RV 
residuals from the best-fit orbit (which may include astrophysical jitter), for other instruments 
this is either an estimate of the precision (not including jitter), or the measured standard 
deviation. We only provide this quantity when applicable. 
}
\ifthenelse{\boolean{emulateapj}}{
    \end{deluxetable*}
}{
    \end{deluxetable}
}

\ifthenelse{\boolean{emulateapj}}{
    \begin{deluxetable*}{lrrrrrr}
}{
    \begin{deluxetable}{lrrrrrr}
}
\tablewidth{0pc}
\tablecaption{
    Relative Radial Velocities and Bisector Spans of \hatcur{}.
    \label{tab:rvs}
}
\tablehead{
    \colhead{BJD} & 
    \colhead{RV\tablenotemark{a}} & 
    \colhead{\ensuremath{\sigma_{\rm RV}}\tablenotemark{b}} & 
    \colhead{BS} & 
    \colhead{\ensuremath{\sigma_{\rm BS}}} & 
        \colhead{Phase} &
        \colhead{Instrument}\\
    \colhead{\hbox{(2\,450\,000$+$)}} & 
    \colhead{(\ms)} & 
    \colhead{(\ms)} &
    \colhead{(\ms)} &
    \colhead{} &
        \colhead{} &
        \colhead{}
}
\startdata
$ 6255.01387 $ & $   129.47 $ & $    56.77 $ & \nodata      & \nodata      & $   0.722 $ & TRES \\
$ 6352.82025 $ & $  -152.51 $ & $    32.59 $ & \nodata      & \nodata      & $   0.276 $ & TRES \\
$ 6576.98851 $ & $   155.31 $ & $    32.59 $ & \nodata      & \nodata      & $   0.808 $ & TRES \\
$ 6585.08850 $ & $  -129.45 $ & $     3.11 $ & \nodata      & \nodata      & $   0.332 $ & HIRES \\
$ 6585.10625 $ & \nodata      & \nodata      & $   -7.8 $ & $   20.9 $ & $   0.340 $ & HIRES \\
$ 6596.70034 $ & $   -18.23 $ & $    42.50 $ & $ 11517.0 $ & $   85.0 $ & $   0.384 $ & Sophie \\
$ 6597.63007 $ & $   112.07 $ & $    22.60 $ & $ 16666.0 $ & $   45.2 $ & $   0.789 $ & Sophie \\
$ 6597.65777 $ & $   142.37 $ & $    15.60 $ & $ 24376.0 $ & $   31.2 $ & $   0.801 $ & Sophie \\
$ 6602.57256 $ & $    72.97 $ & $    15.90 $ & $ 25886.0 $ & $   31.8 $ & $   0.939 $ & Sophie \\
$ 6637.99344 $ & $  -115.18 $ & $     2.68 $ & $    4.9 $ & $   34.7 $ & $   0.350 $ & HIRES \\
$ 6639.12704 $ & $   137.20 $ & $     3.30 $ & $    3.1 $ & $   25.8 $ & $   0.844 $ & HIRES \\
$ 7353.06851 $ & $   -39.00 $ & $     2.42 $ & $    3.2 $ & $   20.3 $ & $   0.468 $ & HIRES \\
$ 7353.93251 $ & $   101.52 $ & $     3.14 $ & $   51.3 $ & $   63.6 $ & $   0.844 $ & HIRES \\
$ 7355.01022 $ & $  -143.22 $ & $     4.11 $ & $  -19.7 $ & $   41.6 $ & $   0.313 $ & HIRES \\

\enddata
\tablenotetext{a}{
        Relative RVs, with $\gamma_{RV}$ subtracted.
}
\tablenotetext{b}{
        Internal errors excluding the component of
        astrophysical/instrumental jitter considered in
        \refsecl{isoparams}.
}
\ifthenelse{\boolean{emulateapj}}{
    \end{deluxetable*}
}{
    \end{deluxetable}
}


\subsection{Photometric Follow-up}
\label{sec:phfu}

\begin{figure}[t]
\epsscale{1.25}
\plotone{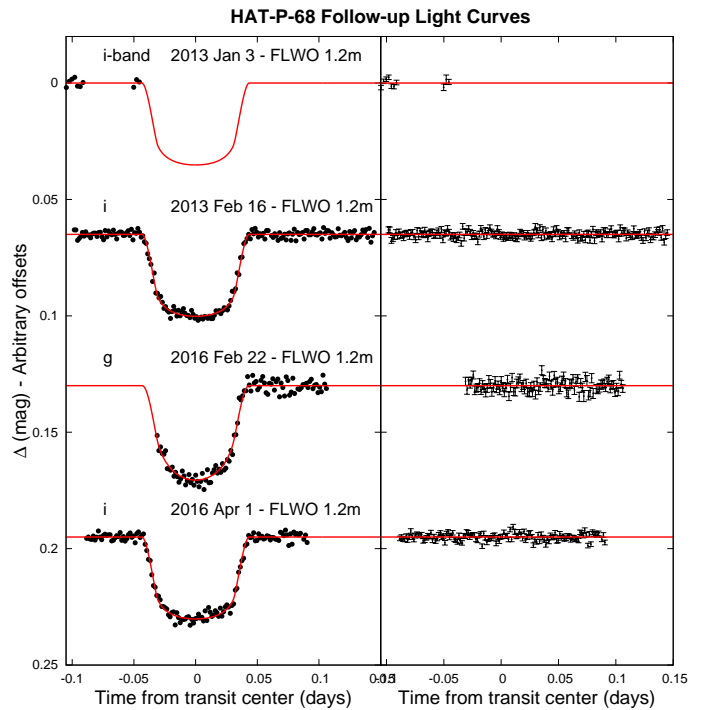}
\caption{
        Unbinned follow-up transit light curves obtained with KeplerCam on the FLWO~1.2\,m, plotted with the best-fit transit model as a solid red line. The dates of observation and photometric filters used are indicated. The residuals are shown on the right-hand side in the same order as the light curves.\\
\label{fig:lc}} 
\end{figure}
In order to confirm the transit signal identified in the HATNet light
curve of \hatcur{}, we carried out photometric follow-up observations of
the system using the KeplerCam mosaic CCD imager on the FLWO~1.2\,m
telescope. Observations used in the analysis were conducted on five
nights covering four predicted primary transit events, and one
predicted secondary eclipse event. The nights, filters, number of
exposures, effective cadences, and point-to-point photometric
precision achieved are listed in Table~\ref{tab:photobs}. A sixth
observation obtained on the night of 2013 Feb 10 did not observe
either the primary transit or secondary eclipse, and was excluded from
the analysis.

The KeplerCam CCD images were calibrated and reduced to light curves
using the aperture photometry routine described by
\citet{bakos:2010:hat11}. We applied an External Parameter
Decorrelation (EPD) and TFA-filtering of the light curves as part of
the global modeling of the system, which we discuss further in
Section~\ref{sec:analysis}. The four light curves covering the primary
transit are shown in Figure~\ref{fig:lc}. The \lc{} covering the
predicted secondary eclipse was consistent with no eclipse variation,
and was used in the blend analysis of the system, but was not included
in the global analysis to determine the planetary and stellar
parameters. All of the \lc{} data are made available in
Table~\ref{tab:phfu}.

\ifthenelse{\boolean{emulateapj}}{
    \begin{deluxetable*}{llrccr}
}{
    \begin{deluxetable}{llrccr}
}
\tablewidth{0pc}
\tablecaption{
    Summary of Photometric Observations
    \label{tab:photobs}
}
\tablehead{
    \multicolumn{1}{c}{Instrument/Field\tablenotemark{a}} &
    \multicolumn{1}{c}{Date(s)} &
    \multicolumn{1}{c}{\# Images} &
    \multicolumn{1}{c}{Cadence\tablenotemark{b}} &
    \multicolumn{1}{c}{Filter} &
    \multicolumn{1}{c}{Precision\tablenotemark{c}} \\
    \multicolumn{1}{c}{} &
    \multicolumn{1}{c}{} &
    \multicolumn{1}{c}{} &
    \multicolumn{1}{c}{(sec)} &
    \multicolumn{1}{c}{} &
    \multicolumn{1}{c}{(mmag)}
}
\startdata
~~~~HAT-5/G268 & 2011 Nov--2012 May & 5867 & 216 & $r$ & 25.6 \\
~~~~HAT-8/G268 & 2012 Jan--2012 Mar & 3034 & 213 & $r$ & 21.8 \\
~~~~FLWO~1.2\,m/KeplerCam & 2013 Jan 03 & 18 & 194 & $i$ & 3.5 \\
~~~~FLWO~1.2\,m/KeplerCam & 2013 Feb 16 & 184 & 114 & $i$ & 1.6 \\
~~~~FLWO~1.2\,m/KeplerCam & 2016 Feb 22 & 102 & 118 & $g$ & 3.3 \\
~~~~FLWO~1.2\,m/KeplerCam & 2016 Mar 24 & 120 & 117 & $i$ & 1.3 \\
~~~~FLWO~1.2\,m/KeplerCam & 2016 Apr 01 & 133 & 118 & $i$ & 1.8 \\
\enddata
\tablenotetext{a}{
    For HATNet data we list the HATNet instrument and field name from
    which the observations are taken. HAT-5 is located at FLWO and
    HAT-8 at MKO. Each field corresponds to one of 838 fixed pointings
    used to cover the full 4$\pi$ celestial sphere. All data from a
    given HATNet field are reduced together, while detrending through 
    External Parameter Decorrelation (EPD) is done independently for 
    each unique unit+field combination.
}
\tablenotetext{b}{
    The median time between consecutive images rounded to the nearest
    second. Due to factors such as weather, the day--night cycle, and
    guiding and focus corrections, the cadence is only approximately
    uniform over short timescales.
}
\tablenotetext{c}{
    The RMS of the residuals from the best-fit model.
} \ifthenelse{\boolean{emulateapj}}{
    \end{deluxetable*}
}{
    \end{deluxetable}
}

\ifthenelse{\boolean{emulateapj}}{
        \begin{deluxetable*}{lrrrcr} }{
        \begin{deluxetable}{lrrrcr} 
    }
\tablewidth{0pc}
\tablecaption{Differential Photometry of \hatcur{} \label{tab:phfu}} 
    \tablehead{ \colhead{BJD} &
    \colhead{Mag\tablenotemark{a}} &
    \colhead{\ensuremath{\sigma_{\rm Mag}}} &
    \colhead{Mag(orig)\tablenotemark{b}} & \colhead{Filter} &
    \colhead{Instrument} \\ 
    \colhead{\hbox{~~~~(2\,400\,000$+$)~~~~}}
    & \colhead{} & \colhead{} & \colhead{} & \colhead{} &
    \colhead{} } 
\startdata 
$ 55912.04050 $ & $  -0.04986 $ & $   0.02247 $ & $ \cdots $ & $ r$ &     HATNet\\
$ 55971.79932 $ & $   0.02491 $ & $   0.01649 $ & $ \cdots $ & $ r$ &     HATNet\\
$ 55932.72681 $ & $  -0.05947 $ & $   0.02526 $ & $ \cdots $ & $ r$ &     HATNet\\
$ 55958.00976 $ & $   0.00106 $ & $   0.01655 $ & $ \cdots $ & $ r$ &     HATNet\\
$ 55948.81621 $ & $  -0.01081 $ & $   0.01803 $ & $ \cdots $ & $ r$ &     HATNet\\
$ 55955.71161 $ & $   0.02603 $ & $   0.01807 $ & $ \cdots $ & $ r$ &     HATNet\\
$ 55994.78471 $ & $  -0.02981 $ & $   0.04192 $ & $ \cdots $ & $ r$ &     HATNet\\
$ 55925.83274 $ & $  -0.02223 $ & $   0.02047 $ & $ \cdots $ & $ r$ &     HATNet\\
$ 55978.69653 $ & $   0.02053 $ & $   0.02025 $ & $ \cdots $ & $ r$ &     HATNet\\
$ 55895.95405 $ & $  -0.00572 $ & $   0.01809 $ & $ \cdots $ & $ r$ &     HATNet\\

\enddata 
\tablenotetext{a}{
     The out-of-transit level has been subtracted. For the HATNet
     light curve, these magnitudes have been detrended using the EPD 
     and TFA procedures prior to fitting a transit model to the light curve. For the
     follow-up light curves derived for instruments other than HATNet,
     these magnitudes have been detrended with the EPD and TFA
     procedure, carried out simultaneously with the transit fit.
}
\tablenotetext{b}{
        Raw magnitude values without application of the EPD and TFA
        procedure. This is only reported for the follow-up light
        curves.
}
\tablecomments{
        This table is available in a machine-readable form in the
        online journal. An abridged version is shown here for guidance
        regarding its form and content. The data are also available on
        the HATNet website at \url{http://www.hatnet.org}.
} \ifthenelse{\boolean{emulateapj}}{ \end{deluxetable*} }{ \end{deluxetable} }


\subsection{Search for Resolved Stellar Companions}
\label{sec:luckyimaging}



\begin{figure}[t]
\centering
\includegraphics*[width=0.475\textwidth]{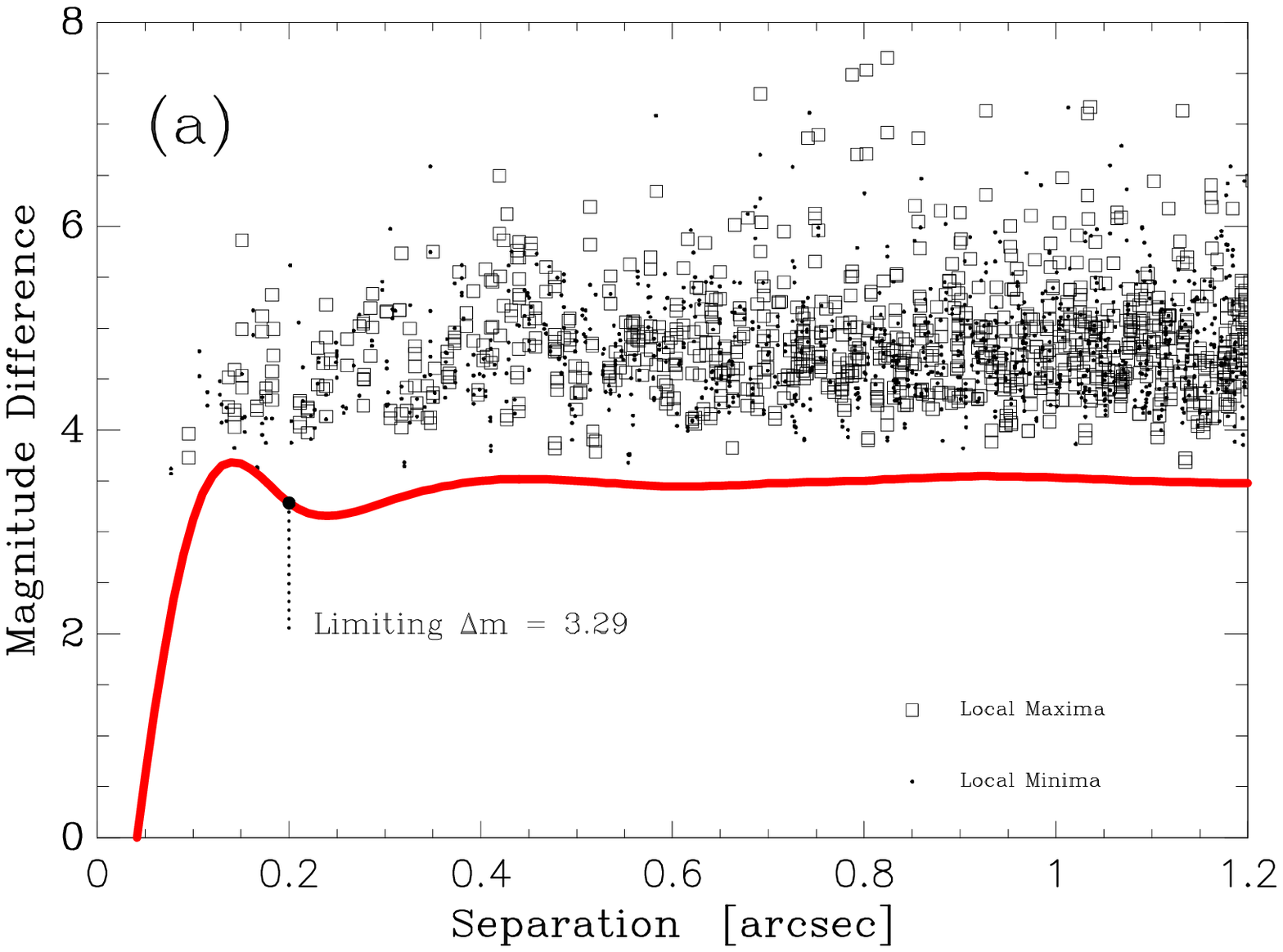}
\includegraphics*[width=0.475\textwidth]{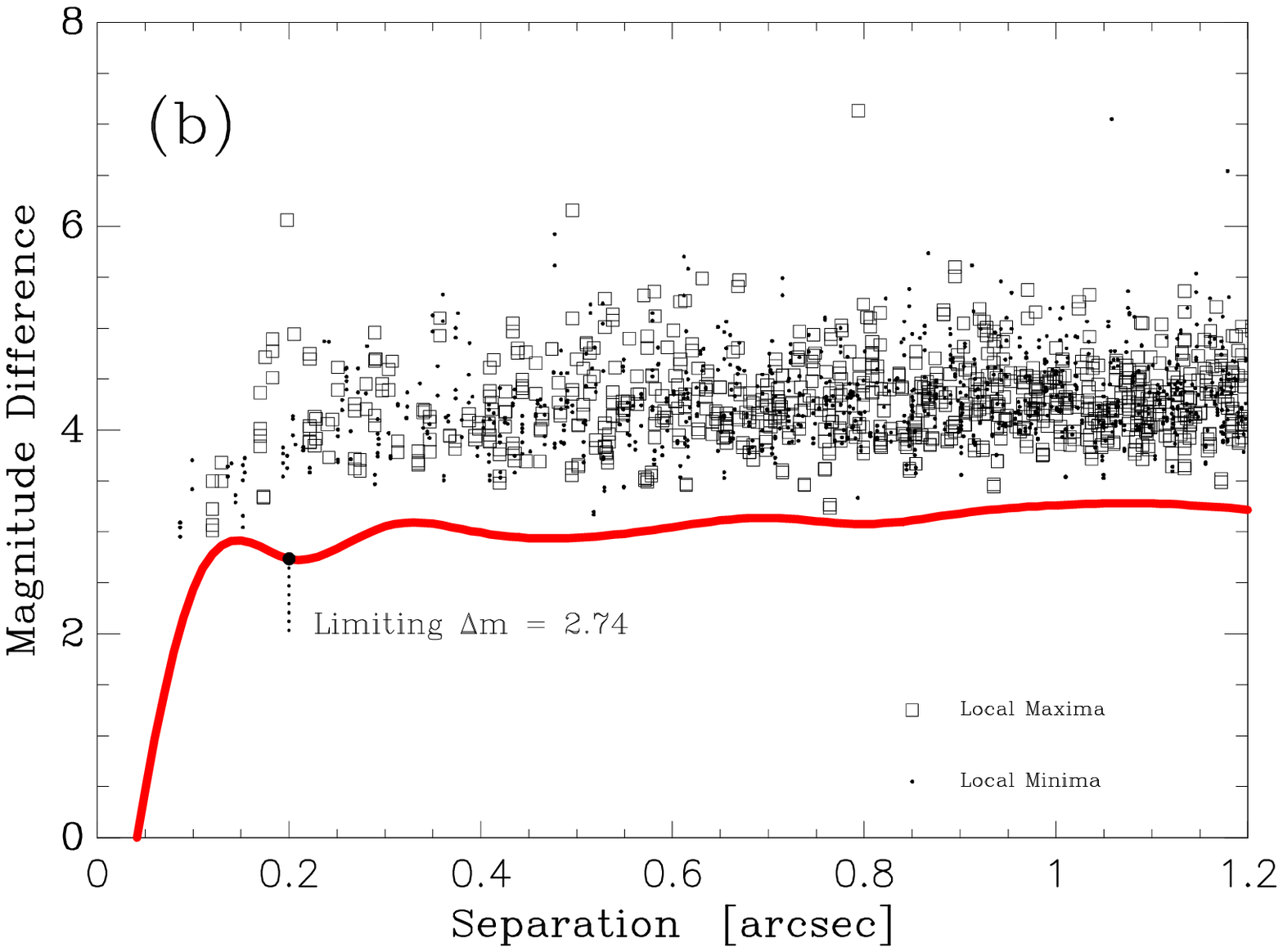}
\caption{
	DSSI detection limits on the relative magnitude of a resolved companion for \hatcur{} in the 692\,nm (a) and 880\,nm (b) filters. The plot shows the local maxima and minima data (in squares and points, respectively) as a function of angular separation. The solid red curves are cubic-spline interpolations of the 5$\sigma$ detection limit.\\
\label{fig:luckyimage}}
\end{figure}
If there are nearby stellar companions to \hatcur{}, they would dilute the transit signal. In order to check for such companions
we obtained high spatial resolution speckle imaging
observations of \hatcur{} with the Differential Speckle Survey
Instrument 
\citep[DSSI;][]{horch:2009} on the
WIYN~3.5\,m telescope\footnote{The WIYN Observatory is a joint
facility of the University of Wisconsin-Madison, Indiana University,
the National Optical Astronomy Observatory and the University of
Missouri.} at Kitt Peak National Observatory in Arizona. 
The observations were gathered on the night of UT 27 October 2015. A
dichroic beamsplitter was used to obtain simultaneous imaging through
692\,nm and 880\,nm filters. 

Each observation consists of a sequence of 1000 40\,ms exposures 
read-out on $128 \times 128$ pixel ($2\farcs8\times 2\farcs8$) subframes, 
that 
are reduced to reconstructed images following \citet{howell:2011}. 
These images were then searched for companions. Finding no companions to 
\hatcur\ within $1\farcs2$ when the ten observations of this system 
were combined, we place $5\sigma$ lower limits on the differential magnitude 
between a putative companion and the primary star as a function of angular separation 
following the method described in \citet{horch:2011}. 

Figure~\ref{fig:luckyimage} shows 
the limiting-magnitude plots constructed from the reconstructed images, 
where the data represent local maxima and minima and the solid curve is a cubic-spline 
interpolation of the 5$\sigma$ detection limit. We find limiting magnitude
differences at $0\farcs2$ of $\Delta m_{692} = 3.29$ and $\Delta m_{880} = 2.74$.

In addition to the companion limits based on the WIYN~3.5\,m/DSSI
observations we also queried the UCAC~4 catalog \citep{zacharias:2013:ucac4}, 
and the \textit{Gaia} DR1 catalog \citep{gaiadr1} for neighbors 
within 20\arcsec\ that may dilute either the HATNet or KeplerCam 
photometry. We find no such neighbors. Additionally, the \textit{Gaia} DR2 
catalog \citep{gaiadr2} shows no neighbors within 10\arcsec\ of \hatcur{}.

\section{Analysis}
\label{sec:analysis}

We analyzed the photometric and spectroscopic observations of
\hatcur{} to determine the parameters of the system using the 
most up-to-date
procedures developed for HATNet and HATSouth \citep{hartman:2019:hats69,bakos:2018:hats71}.
In the following, we 
briefly summarize
our analysis methods to accurately 
determine the stellar and planetary physical parameters and 
to rule out various false positive scenarios.

\subsection{Stellar Host Properties}
\label{sec:starparams}

High-precision stellar atmospheric parameters were measured from the
I$_{2}$-free HIRES template spectrum using SPC, yielding $\teffstar =
\hatcurSMEteff$\,K, \feh$=\hatcurSMEzfeh$,
$\vsini=\hatcurSMEvsin$\,\kms, and $\log g_{\star} = \hatcurSMElogg$
(cgs). The resulting \teffstar\ and \feh\ measurements were 
included in the global modeling to determine the physical stellar parameters.

We ultimately tried three methods to ascertain these physical parameters.
The first two methods compare the observable properties to two different stellar evolution models.
The last method uses empirical relations to derive stellar mass and radius.

\subsubsection{Isochrone-based Parameters}
\label{sec:isoparams}

Initially, we attempted to compare the Yonsei-Yale \citep[Y$^{2}$;][]{yi:2001} 
models to the observed light-curve-based stellar density, 
and the spectroscopically determined values of \teffstar\ and \feh. This is the method that was followed, for example, in \citet{bakos:2010:hat11}, and has been previously applied to the majority of published transiting planet discoveries from the HATNet project.
Note that this was completed prior to the availability of \textit{Gaia} DR2. 
Assuming a circular orbit, the best-fit stellar density is more than 
$3\sigma$ lower than the minimum density from theoretical models
-- that was achieved within the age of the Galaxy for a K dwarf star with a photosphere temperature 
of $4500$\,K. This discrepancy between the measured stellar density 
and 
older
stellar evolution models,
such as the Y$^{2}$ models,
has been previously 
reported
for other mid K through early M dwarf stars \citep[eg.][]{boyajian:2012}. 

Fortunately, \citet{chen:2014:parsec} improved the PAdova-TRieste Stellar Evolution Code 
\citep[PARSEC;][]{bressan:2012:parsec} models for 
very low mass stars ($< 0.6 \msun$) over a wide range of wavelengths. 
\citet{randich:2018} also demonstrated that \textit{Gaia} parallaxes 
can be combined with ground-based datasets to yield robust stellar ages.
As such, we opted to use PARSEC models combined with the \textit{Gaia} DR2 data, following \citet{hartman:2019:hats69}.

We performed a tri-linear interpolation within a grid of PARSEC model isochrones using \teffstar, \feh, and the bulk stellar density \rhostar\ as the independent variables. These three variables in turn are directly varied in the global MCMC analysis (Section~\ref{sec:jointfit}), or determined directly from parameters that are varied in this fit. The tri-linear interpolation then yields the \mstar, \rstar\, \lstar, and age values to associate with each trial set of \teffstar, \feh\ and \rhostar. Through this process we restrict the fit to consider only combinations of \teffstar, \feh\ and \rhostar\ that match to a stellar model. For K dwarf stars, such as \hatcur{}, which exhibit little evolution over the age of the Galaxy, this is a rather restrictive constraint. Including this constraint yields a posteriori estimates for the stellar atmospheric parameters of:
$\teffstar =
\hatcurISOteffisochronecirc$\,K, \feh$=\hatcurISOzfehisochronecirc$,
 and $\log g_{\star} = \hatcurISOloggisochronecirc$. 
Assuming a circular orbit, the PARSEC isochrone-based method yields
a stellar mass and radius of 
\hatcurISOmisochronecirc\,\msun\ and \hatcurISOrisochronecirc\,\rsun, respectively, 
an age of \hatcurISOageisochronecirc\,Gyr, and a 
reddening-corrected distance of \hatcurXdistredisochronecirc\,pc. 


\subsubsection{Empirically Based Parameters}
\label{sec:empparams}


As an alternative approach, we also determined the stellar physical parameters following an empirical method similar to that of \citet{stassun:2017}. This method effectively combines the bulk stellar density 
-- measured from the transit light curve -- with the stellar radius --measured from the effective temperature, parallax and apparent magnitudes in several band-passes -- to determine the stellar mass. In practice this is incorporated into the global MCMC modeling (Section~\ref{sec:jointfit}), and theoretical bolometric corrections are used to predict the absolute magnitude in each band-pass from the effective temperature, radius and metallicity of the star.
Assuming a circular orbit, this empirical method yields a stellar mass and radius of 
\hatcurISOmnoisorestrictempiricalcirc\,\msun\ and \hatcurISOrnoisorestrictempiricalcirc\,\rsun,  
respectively, and a reddening-corrected distance of \hatcurXdistrednoisorestrictempiricalcirc\,pc.
Note that these parameters are not restricted by the isochrones from PARSEC, which is why the uncertainties are larger compared to the uncertainties on the isochrone-based parameters.


\subsection{Global Modeling}
\label{sec:jointfit}

We determined the parameters of the system by carrying out a joint modeling of the
high-precision RVs (fit using a Keplerian orbit), the HATNet and
follow-up light curves (fit using a \citealp{mandel:2002} transit
model with Gaussian priors for the quadratic limb darkening coefficients taken from
\citealp{claret:2012,claret:2013} and \citealp{claret:2018} to place Gaussian prior constraints on their values, assuming a prior uncertainty of $0.2$ for each coefficient), the catalog broad-band magnitudes, the stellar parallax from Gaia~DR2, and the spectroscopically determined atmospheric parameters of the system. These latter stellar observations were modeled using isochrone and empirical-based methods, as discussed above (Section~\ref{sec:starparams}). 
This analysis makes use of a
differential evolution Markov Chain Monte Carlo procedure
\citep[MCMC;][]{terbraak:2006} to estimate the posterior parameter
distributions, which we use to determine the median parameter values
and their 1$\sigma$ uncertainties. 

For each of the methods that we adopted 
to model 
the stellar parameters, we carried out two fits, 
one where the orbit is assumed to be circular, and another where the
eccentricity parameters are allowed to vary in the fit. In both cases
we allow the RV jitter (an extra term added in quadrature to the
formal RV uncertainties) to vary independently for each of the instruments used. 
We find that when the isochrone-based stellar parameters are used, the free eccentricity 
model yields an eccentricity consistent with zero 
($e = \hatcurRVeccenisochroneeccen{}$), resulting in a 95\% confidence upper limit on 
the eccentricity of $e\hatcurRVeccentwosiglimisochroneeccen{}$. We therefore adopt 
the following
parameters for \hatcurb{} assuming a circular orbit: 
a mass of \hatcurPPmlong\,\mjup, a radius of
\hatcurPPrlong\,\rjup, and an equilibrium temperature of $\hatcurPPteff{}$\,K.
The equilibrium temperature
was calculated assuming zero albedo and full redistribution of heat. 
We give the planetary parameters derived from the joint fit, in Table~\ref{tab:planetparam}. 
For comparison, when the empirical method is used, and a circular orbit is assumed, we find a planet mass of \hatcurPPmlongempiricalcirc\,\mjup, planet radius of \hatcurPPrlongempiricalcirc\,\rjup, and an equilibrium temperature of $\hatcurPPteffempiricalcirc{}$\,K.

\subsection{Adopted Parameters}
\label{sec:adopted}

\begin{figure}
\epsscale{1.25}
\plotone{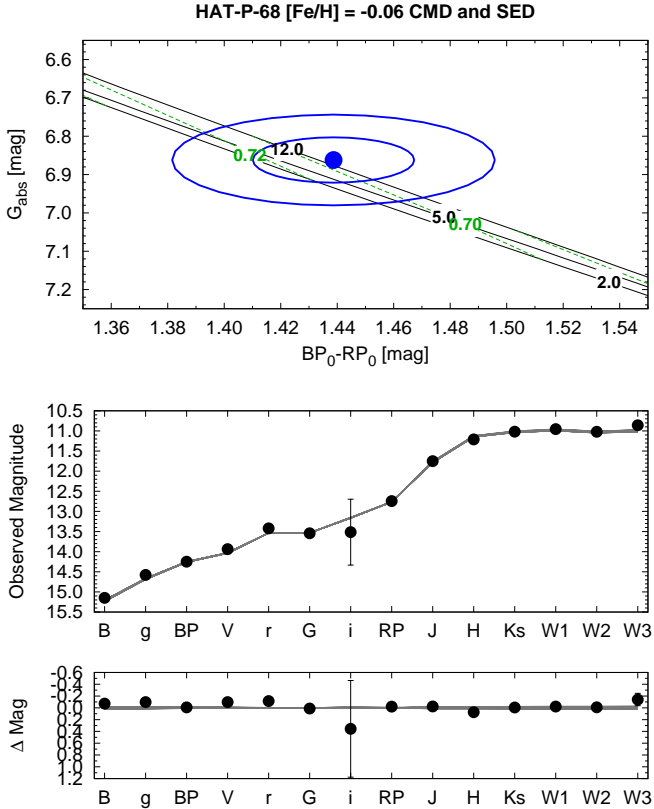}
\caption{
			{\em Top:} The absolute \textit{Gaia} G-band magnitude vs. the dereddened $BP-RP$ color. This measured value is compared to theoretical isochrones (black lines at Gyr ages in black) and stellar evolution tracks (green lines at solar masses in green) from the PARSEC models interpolated at the spectroscopically determined metallicity of the host. The filled blue circle show the measured reddening- and distance-corrected values from \textit{Gaia} DR2, while the blue lines indicate the $1\sigma$ and $2\sigma$ confidence regions, including the estimated systematic errors in the photometry. {\em Middle:} The SED as measured via broadband photometry through the fourteen listed filters. We plot the observed magnitudes without correcting for distance or extinction. Over-plotted are 200 model SEDs randomly selected from the MCMC posterior distribution produced through the global analysis. {\em Bottom:} The residuals from the best-fit model SED.\\
\label{fig:iso}}
\end{figure}
We included in the 
global modeling  
analysis broad-band photometry from \textit{Gaia} DR2, APASS \citep{henden:2009}, 2MASS \citep{skrutskie:2006}, 
and WISE \citep{wright:2010} -- $G$, $BP$, $RP$, $B$, $V$, $g$, $r$, $i$,
$J$, $H$, $K_s$, $W1$, $W2$, and $W3$ bands. 
To account for dust extinction we included $A_{V}$ as a free-parameter in the model, assumed the \citet{cardelli:1989} $R_{V} = 3.1$ extinction law, and placed a Gaussian prior on $A_{V}$ based on the predicted extinction from the MWDUST 3D Galactic extinction model \citep{bovy:2016}.

Figure~\ref{fig:iso} shows the comparison between the 
broadband photometric measurements and the PARSEC models mentioned 
in Section~\ref{sec:isoparams}. 
The top panel is a \cmd{} 
(CMD) 
of the \textit{Gaia} $G$ magnitude versus the 
dereddened $BP - RP$ color as a filled blue circle, along with the 
$1\sigma$ and $2\sigma$ confidence regions in blue lines.
We plot a set of $\feh$=$0.06$ isochrones 
and stellar evolution tracks using black lines and green lines, respectively.
The age of each isochrone is listed in black using Gyr units, while the mass of each evolution track 
is listed in green using solar mass units. The middle panel compares 200 model 
spectral energy distributions (SEDs) to the observed broadband photometry, 
the latter of which has not been corrected for distance or extinction. 
The bottom panel shows the residuals from the best-fit model SED.
We find that the observed photometry and parallax is consistent with the models.

Based on the global MCMC analysis, 
we adopt parameters assuming a circular orbit for the isochrone-based method. 
We list the joint fit derived parameters in Tables~\ref{tab:stellar} and ~\ref{tab:planetparam}. 

\subsection{Excluding False Positive Scenarios}
\label{sec:blend}

In order to rule out the possibility that \hatcur{} is a blended
stellar eclipsing binary 
(EB) 
system, we carried out a direct blend analysis of
the data following \citet{hartman:2012:hat39hat41}, 
with modifications from \citet{hartman:2019:hats69}. We
find that all blended stellar 
EB 
models tested can be ruled out -- based on their fit to the
photometry, parallax, and light curves -- with almost $4\sigma$ confidence, and conclude
that \hatcur{} is a transiting planet system, and not a blended
stellar 
EB 
system. 

Note that the blend analysis of \hatcur{} as an unresolved stellar binary with a planet around one 
stellar component provides a slight improvement to the fit compared to assuming no such unresolved stellar companion ($\Delta \chi^{2}$ value of $-2.93$), but the difference is consistent with the expected improvement from adding an additional parameter to the fit. 
Based on the high-spatial-resolution imaging that we have carried out (Section~\ref{sec:luckyimaging}), any unresolved companion separated by more than $\sim 0\farcs2$, must have $\Delta m > 3.07$ at 692\,nm compared to the transiting planet host. 
We conclude our findings 
assuming that there is no stellar companion.

\ifthenelse{\boolean{emulateapj}}{
  \begin{deluxetable*}{lcr}
}{
  \begin{deluxetable}{lcr}
}
\tablewidth{0pc}
\tablecaption{
    Stellar Parameters for \hatcur{} 
    \label{tab:stellar}
}
\tablehead{
    \multicolumn{1}{c}{~~~~~~~~Parameter~~~~~~~~} &
    \multicolumn{1}{c}{Value}                     &
    \multicolumn{1}{r}{Source}    
}
\startdata
\noalign{\vskip -3pt}
\sidehead{Identifiers}
~~~~GSC-ID  &   \hatcurCCgsc{} & \\
~~~~2MASS-ID   &  \hatcurCCtwomass{} & \\
~~~~\textit{Gaia} DR2-ID   & \hatcurCCgaiadrtwo{} & \\
\sidehead{Astrometric Properties}
~~~~R.A.~(h:m:s)                      &  \hatcurCCra{} & \textit{Gaia} DR2\\
~~~~Dec.~(d:m:s)                      &  \hatcurCCdec{} & \textit{Gaia} DR2\\
~~~~R.A.p.m.~(mas/yr)                 &  \hatcurCCpmra{} & \textit{Gaia} DR2\\
~~~~Dec.p.m.~(mas/yr)                 &  \hatcurCCpmdec{} & \textit{Gaia} DR2\\
~~~~Parallax~(mas)           		& \hatcurCCparallax{} & \textit{Gaia} DR2\\
\sidehead{Spectroscopic Properties}
~~~~$\teffstar$ (K)\dotfill         &  \hatcurSMEteff{} & SPC\tablenotemark{a}    \\
~~~~$\feh$\dotfill                  &  \hatcurSMEzfeh{} & SPC                 \\
~~~~$\vsini_{\star}$ (\kms)\dotfill   &  \hatcurSMEvsin{} & SPC                 \\
\sidehead{Photometric Properties}
~~~~$G$ (mag)\tablenotemark{b}\dotfill               & \hatcurCCgaiamG{} & \textit{Gaia} DR2    \\
~~~~$BP$ (mag)\tablenotemark{b}\dotfill 				& \hatcurCCgaiamBP{} & \textit{Gaia} DR2 	\\
~~~~$RP$ (mag)\tablenotemark{b}\dotfill				& \hatcurCCgaiamRP{} & \textit{Gaia} DR2 	\\
~~~~$B$ (mag)\dotfill               &  \hatcurCCtassmB{} & APASS                \\
~~~~$V$ (mag)\dotfill               &  \hatcurCCtassmv{} & APASS               \\
~~~~$g$ (mag)\dotfill               &  \hatcurCCtassmg{} & APASS                \\
~~~~$r$ (mag)\dotfill               &  \hatcurCCtassmr{} & APASS                \\
~~~~$i$ (mag)\dotfill               &  \hatcurCCtassmi{} & APASS                \\
~~~~$J$ (mag)\dotfill               &  \hatcurCCtwomassJmag{} & 2MASS           \\
~~~~$H$ (mag)\dotfill               &  \hatcurCCtwomassHmag{} & 2MASS           \\
~~~~$K_s$ (mag)\dotfill             &  \hatcurCCtwomassKmag{} & 2MASS           \\
~~~~$W1$ (mag)\dotfill             &  \hatcurCCWonemag{} & WISE           \\
~~~~$W2$ (mag)\dotfill             &  \hatcurCCWtwomag{} & WISE           \\
~~~~$W3$ (mag)\dotfill             &  \hatcurCCWthreemag{} & WISE           \\
~~~~$P_{\rm rot}$ (days)\dotfill    &  $24.593 \pm 0.064$ & HATNet           \\
\sidehead{Derived Properties} 
~~~~$\mstar$ ($\msun$)\dotfill      &  \hatcurISOmlong{} & Global Modeling \tablenotemark{c}\\
~~~~$\rstar$ ($\rsun$)\dotfill      &  \hatcurISOrlong{} & Global Modeling \\
~~~~$\loggstar$ (cgs)\dotfill       &  \hatcurISOlogg{} & Global Modeling \\
~~~~$\rhostar$ (\gcmc) \dotfill 	& \hatcurISOrho{} & Global Modeling \\
~~~~$\lstar$ ($\lsun$)\dotfill      &  \hatcurISOlum{} & Global Modeling \\ 
~~~~$\teffstar$ (K) \dotfill 		& \hatcurISOteff{} & Global Modeling \\
~~~~$\feh$\dotfill                  & \hatcurISOzfeh{} & Global Modeling \\
~~~~Age (Gyr)\dotfill               &  \hatcurISOage{} & Global Modeling \\
~~~~$A_{V}$ (mag) \dotfill           &  \hatcurXAv{} & Global Modeling \\
~~~~Distance (pc)\dotfill           &  \hatcurXdistred{} & Global Modeling \\
\enddata
\tablenotetext{a}{
    SPC = ``Stellar Parameter Classification'' method
    for the analysis of high-resolution spectra \citep{buchhave:2012:spc}
    applied to the Keck-HIRES I$_{2}$-free template spectrum of \hatcur{}. }
\tablenotetext{b}{The listed uncertainties for the \textit{Gaia} DR2 photometry are taken from the catalog. 
For the analysis we assume additional systematic uncertainties of $0.002$, $0.005$, and $0.003$ mag for 
the $G$, $BP$, and $RP$ bands, respectively.}
\tablenotetext{c}{
   A posteriori estimates from the Global MCMC analysis of the observations described in Section~\ref{sec:jointfit}. The parameters presented here are derived from an analysis where the stellar parameters are constrained using the PARSEC stellar evolution models \citep{bressan:2012:parsec}, and a circular orbit is assumed for the planet.
}
\ifthenelse{\boolean{emulateapj}}{
  \end{deluxetable*}
}{
  \end{deluxetable}
}


\ifthenelse{\boolean{emulateapj}}{
  \begin{deluxetable*}{lrlr}
}{
  \begin{deluxetable}{lrlr}
}
\tablecolumns{4}
\tablecaption{Parameters for the planet \hatcurb{}.
\label{tab:planetparam}}
\tablehead{
    \multicolumn{1}{c}{~~~~~~~~Parameter~~~~~~~~} &
    \multicolumn{1}{r}{Value \tablenotemark{a}} &
    \multicolumn{1}{c}{~~~~~~~~Parameter~~~~~~~~} &
    \multicolumn{1}{r}{Value \tablenotemark{a}}
}
\startdata
\noalign{\vskip -2pt}
\multicolumn{2}{l}{{\em \Lc{} parameters}} & \multicolumn{2}{l}{{\em RV parameters}} \\
~~~$P$ (days)             \dotfill    & $\hatcurLCP{}$ & ~~~$K$ (\ms)              \dotfill    & $\hatcurRVK{}$ \\
~~~$T_c$ (${\rm BJD}$)    
      \tablenotemark{b}   \dotfill    & $\hatcurLCT{}$ & ~~~$e$ \tablenotemark{e}  \dotfill    & $\hatcurRVeccentwosiglimisochroneeccen{}$ \\
~~~$T_{14}$ (days)
      \tablenotemark{b}   \dotfill    & $\hatcurLCdur{}$ & ~~~HIRES RV jitter (\ms) \tablenotemark{f}        \dotfill    & \hatcurRVjitterA{} \\
~~~$T_{12} = T_{34}$ (days)
      \tablenotemark{b}   \dotfill    & $\hatcurLCingdur{}$ & ~~~TRES RV jitter (\ms) \tablenotemark{f}        \dotfill    & \hatcurRVjitterB{} \\
~~~$\arstar$              \dotfill    & $\hatcurPPar{}$ & ~~~Sophie RV jitter (\ms) \tablenotemark{f}        \dotfill    & \hatcurRVjitterC{} \\
~~~$\zrstar$ \tablenotemark{c}              \dotfill    & $\hatcurLCzeta{}$  & \multicolumn{2}{l}{{\em Planetary parameters}}
\\
~~~$\rpl/\rstar$          \dotfill    & $\hatcurLCrprstar{}$  & ~~~$\mpl$ ($\mjup$)       \dotfill    & $\hatcurPPmlong{}$ \\
~~~$b^2$                  \dotfill    & $\hatcurLCbsq{}$  & ~~~$\rpl$ ($\rjup$)       \dotfill    & $\hatcurPPrlong{}$ \\
~~~$b \equiv a \cos i/\rstar$
                          \dotfill    & $\hatcurLCimp{}$ & ~~~$C(\mpl,\rpl)$
    \tablenotemark{g}     \dotfill    & $\hatcurPPmrcorr{}$ \\
~~~$i$ (deg)              \dotfill    & $\hatcurPPi{}$ & ~~~$\rhopl$ (\gcmc)       \dotfill    & $\hatcurPPrho{}$ \\
\multicolumn{2}{l}{{\em Limb-darkening coefficients} \tablenotemark{d}} & ~~~$\log g_p$ (cgs)       \dotfill    & $\hatcurPPlogg{}$ \\
~~~$c_1,g$ (linear term) \dotfill    & $\hatcurLBig{}$    & ~~~$a$ (AU)               \dotfill    & $\hatcurPParel{}$ \\
~~~$c_2,g$ (quadratic term) \dotfill    & $\hatcurLBiig{}$   & ~~~$T_{\rm eq}$ (K) \tablenotemark{h}        \dotfill   & $\hatcurPPteff{}$  \\
~~~$c_1,r$              \dotfill    & $\hatcurLBir{}$    & ~~~$\Theta$ \tablenotemark{i} \dotfill & $\hatcurPPtheta{}$  \\
~~~$c_2,r$               \dotfill    & $\hatcurLBiir{}$   & ~~~$\langle F \rangle$ (\ergscmsq) \tablenotemark{j}
                          \dotfill    & $(\hatcurPPfluxavg{}) \times 10^{\hatcurPPfluxavgdim}$ \\
~~~$c_1,i$ \dotfill    & $\hatcurLBii{}$   & & \\
~~~$c_2,i$ \dotfill  & $\hatcurLBiii{}$  & & \\
\enddata
\tablenotetext{a}{
    For each parameter we give the median value and
    68.3\% (1$\sigma$) confidence intervals from the posterior
    distribution. Reported results assume a circular orbit.
}
\tablenotetext{b}{
    Reported times are in Barycentric Julian Date calculated directly
    from UTC, {\em without} correction for leap seconds.
    \ensuremath{T_c}: Reference epoch of mid transit that
    minimizes the correlation with the orbital period. 
    \ensuremath{T_{14}}: total transit duration, time
    between first to last contact; 
    \ensuremath{T_{12}=T_{34}}: ingress/egress time, time between first
    and second, or third and fourth contact.
}
\tablenotetext{c}{
    Reciprocal of the half duration of the transit used as a jump
    parameter in our DE-MC analysis in place of $\arstar$. It is
    related to $\arstar$ by the expression $\zrstar = \arstar
    (2\pi(1+e\sin \omega))/(P \sqrt{1 - b^{2}}\sqrt{1-e^{2}})$
    \citep{bakos:2010:hat11}.
}
\tablenotetext{d}{
    Values for a quadratic law, adopted from the tabulations by
    \cite{claret:2004} according to the spectroscopic (SPC) parameters
    listed in \reftabl{stellar}.
}
\tablenotetext{e}{
    The 95\% confidence upper-limit on the eccentricity. All other
    parameters listed are determined assuming a circular orbit for this planet.
}
\tablenotetext{f}{
    Error term, either astrophysical or instrumental in origin, added
    in quadrature to the formal RV errors. This term is varied in the
    fit independently for each instrument assuming a prior that is inversely proportional to the jitter.
}
\tablenotetext{g}{
    Correlation coefficient between the planetary mass \mpl\ and
    radius \rpl\ determined from the parameter posterior distribution
    via $C(\mpl,\rpl) = \langle(\mpl - \langle\mpl\rangle)(\rpl -
    \langle\rpl\rangle)\rangle/(\sigma_{\mpl}\sigma_{\rpl})\rangle$, 
	where $\langle \cdot \rangle$ is the
    expectation value, and $\sigma_x$ is the std.\
    dev.\ of $x$.
}
\tablenotetext{h}{
    Planet equilibrium temperature averaged over the orbit, calculated
    assuming a Bond albedo of zero, and that flux is reradiated from
    the full planet surface.
}
\tablenotetext{i}{
    The Safronov number is given by $\Theta = \frac{1}{2}(V_{\rm
    esc}/V_{\rm orb})^2 = (a/\rpl)(\mpl / \mstar )$
    \citep[see][]{hansen:2007}.
}
\tablenotetext{j}{
    Incoming flux per unit surface area, averaged over the orbit.
}
\ifthenelse{\boolean{emulateapj}}{
  \end{deluxetable*}
}{
  \end{deluxetable}
}
%

\section{Discussion}
\label{sec:discussion}

\begin{figure}[t]
\epsscale{1.25}
\plotone{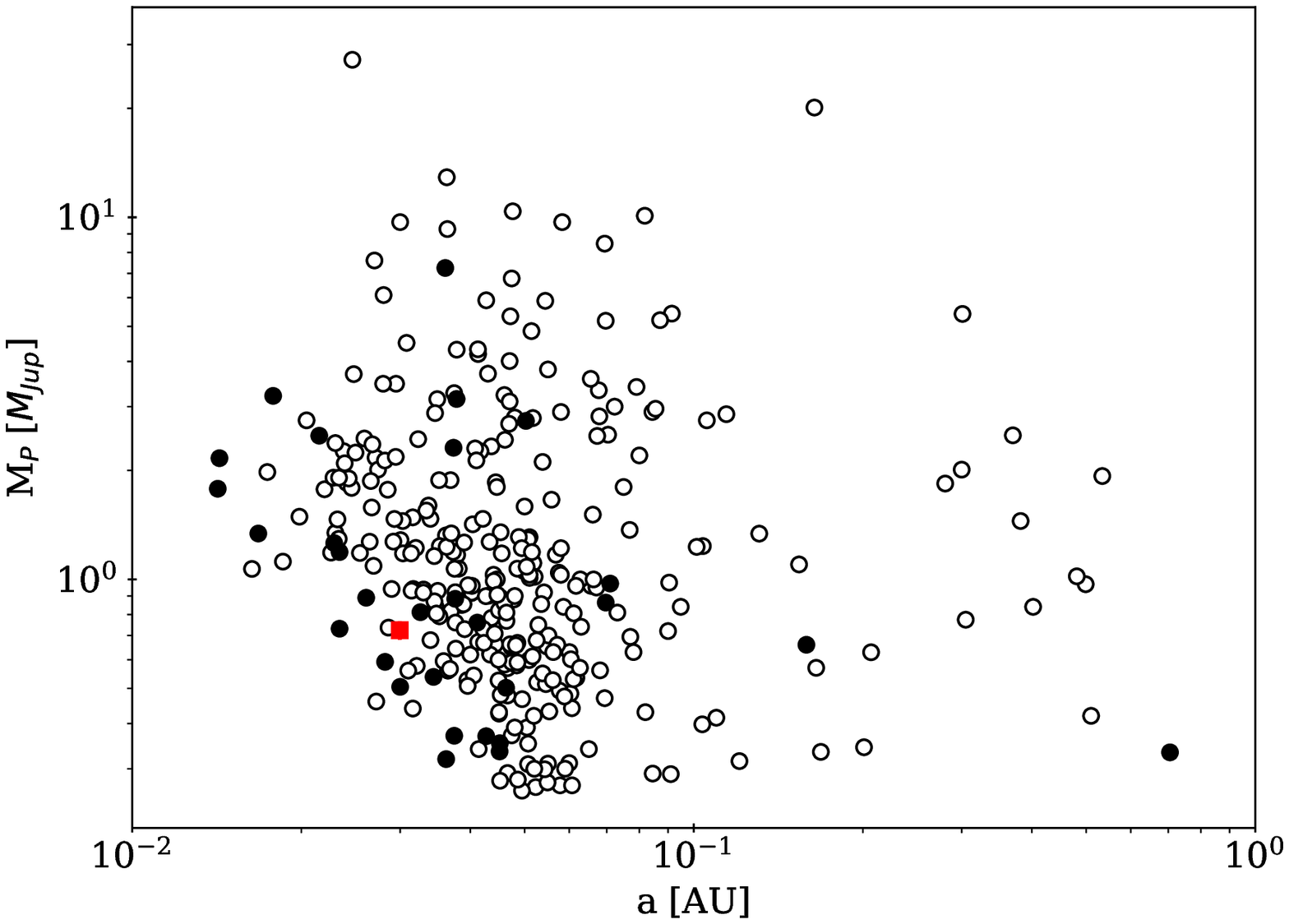}
\caption{
			Planet mass as a function of semi-major axis for all known giant ($M_{P} > 0.25 \mjup$) planets with measured masses and radii. We differentiate host star masses by plotting the data for stars with $< 0.8 \msun$ (as filled black circles), for stars with $> 0.8 \msun$ (as unfilled black circles), and for \hatcurb{} (red square with errorbars). Data from NASA Exoplanet Archive as of 2020 September 24.\\
\label{fig:archive}}
\end{figure}
In this paper we have presented the discovery of the \hatcur{} transiting planet 
system by the HATNet survey. We have found that every $\hatcurLCPshort{}$\,days,
the planet \hatcurb{} -- with a mass of \hatcurPPmlong{}\,\mjup, and radius of \hatcurPPrlong{}\,\rjup, 
-- orbits a star of mass \hatcurISOm{}\,\msun, and radius \hatcurISOr{}\,\rsun. As such, 
the discovery of this planet contributes to the relatively small sample of 
low-mass (late K dwarf, and M dwarf)
stars with known transiting giant planets.

We compared the newly discovered planet to the previously discovered planets listed 
in the NASA Exoplanet Archive as of 2020 September 24. With a semi-major axis of 
$a$=$\hatcurPParel{}$\,AU, this planet joins the 
small but growing sample of 28 known giant planets  
in sub-0.05 AU orbits around low mass stars ($< 0.8 \msun$). 
Here, we
restrict
the sample of confirmed planets to those with well-measured masses greater than $0.25 \mjup$,
following \citet{dawson:2018}. To demonstrate the significance of this planet, we show \hatcurb{} (red square with errorbars) in context with these HJs in Figure~\ref{fig:archive}. The data for stars less massive than $0.8 \msun$ are filled black circles, while data for stars more massive than $0.8 \msun$ are unfilled black circles.


\begin{figure}[t]
\epsscale{1.25}
\plotone{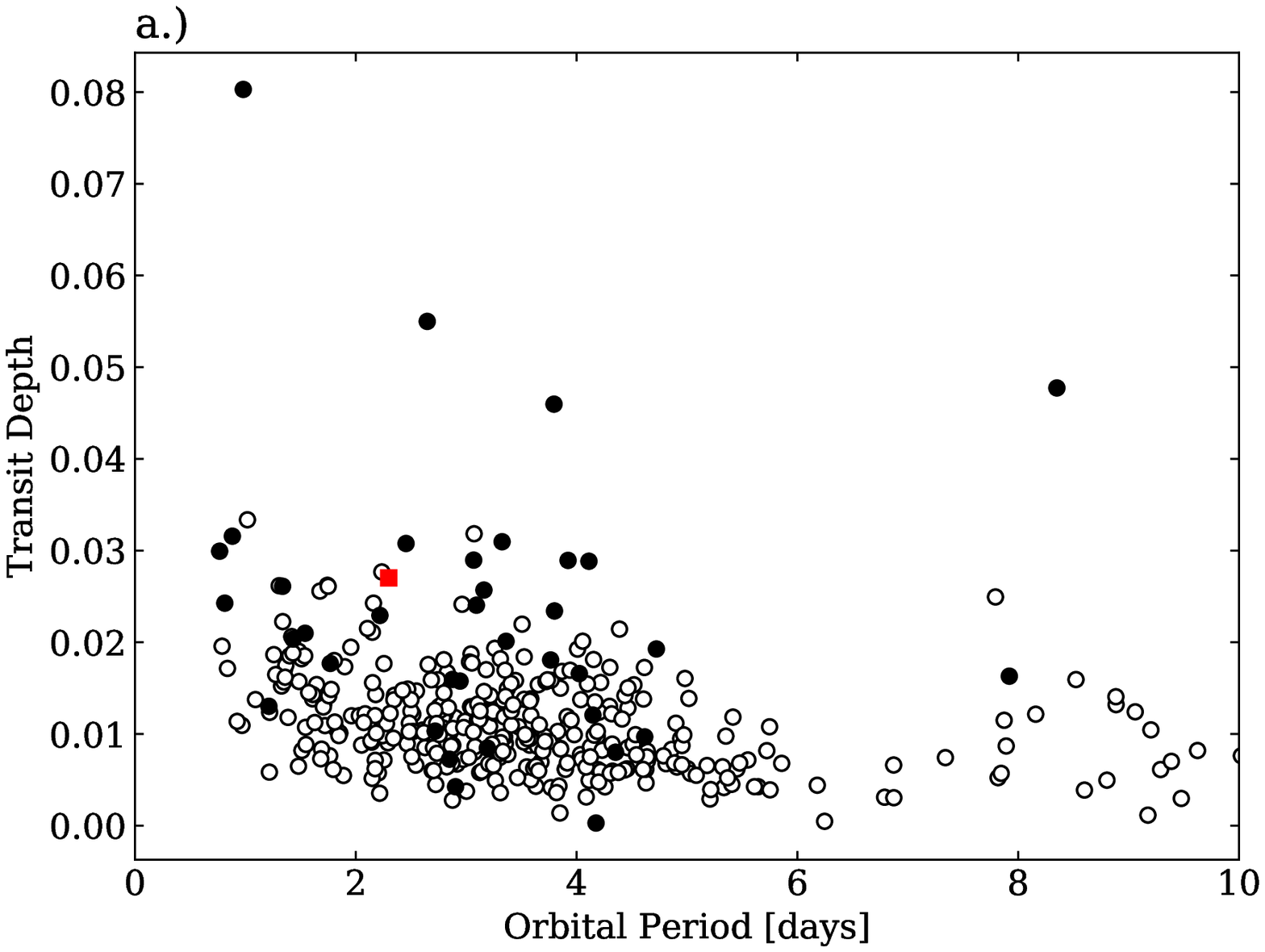}
\plotone{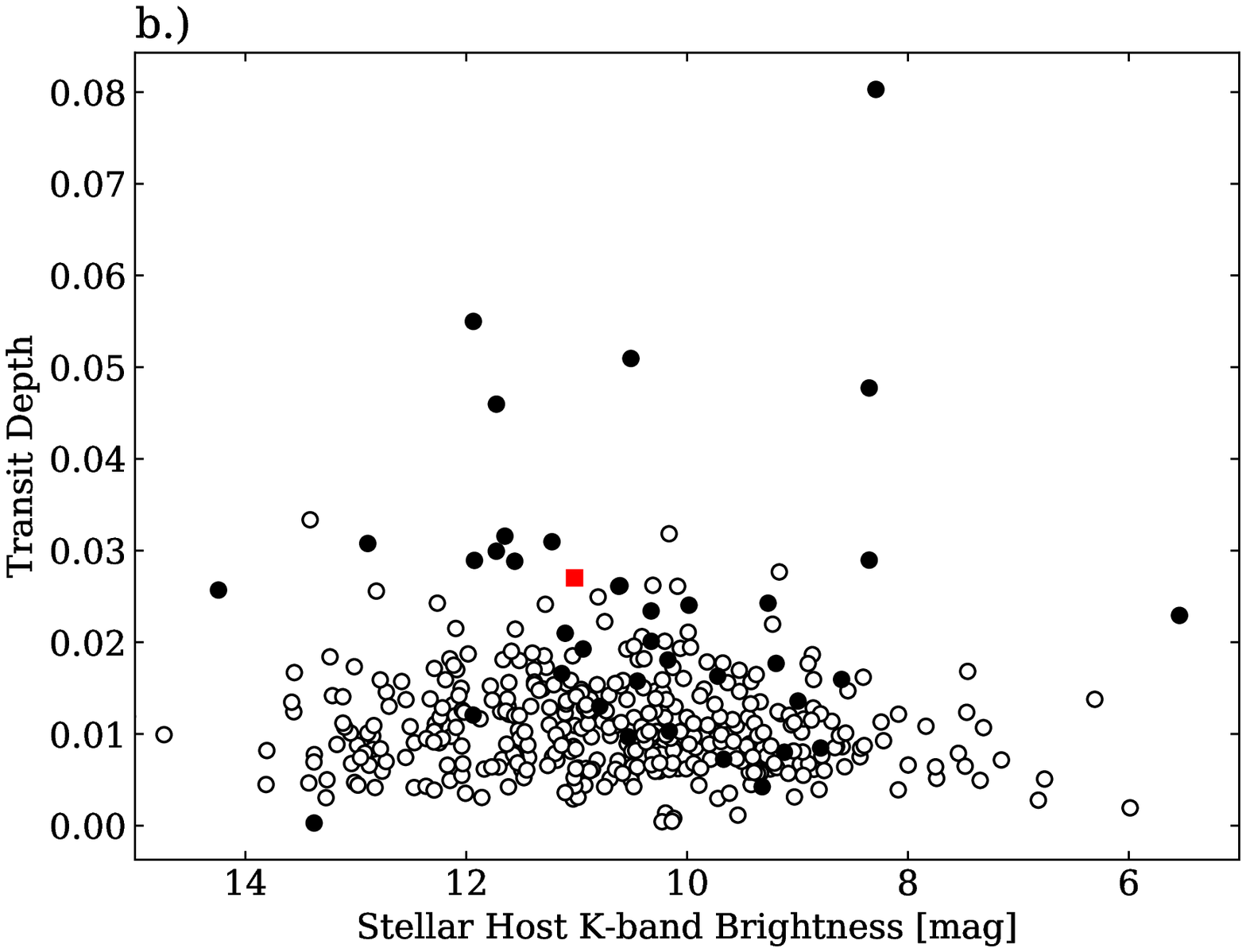}
\caption{
			Transit depth as a function of (a.) orbital period for hot Jupiters and (b.) host K-band magnitude. The depth was calculated from the planetary radius $R_{P}$ and stellar radius $R_{\star}$. We differentiate host star masses by plotting the data for stars with $< 0.8 \msun$ (as filled black circles), for stars with $> 0.8 \msun$ (as unfilled black circles), and for \hatcurb{} (red square with errorbars). Data from NASA Exoplanet Archive as of 2020 September 24.\\
\label{fig:transdepth}}
\end{figure}

We find that including \hatcurb{}, there are 11 planetary systems with transit depths $> 2.5\%$, which may be good targets for transmission spectroscopy. 
Of these other worlds, those that have already been studied using transmission spectroscopy include
WASP-80b \citep{mancini:2014:wasp80,kirk:2017:wasp80}, WASP-52b 
\citep{kirk:2016:wasp52,louden:2017:wasp52} and WASP-43b 
\citep{chen:2014:wasp43,weaver:2020:wasp43}. 
While \hatcur{} is much fainter than these hosts in the optical band-pass, it is only 1\,mag fainter than WASP-52 in the $K$-band. In figure \ref{fig:transdepth}, we plot show the transit depths of HJs as a function of period (a.) and as a function of K-band stellar brightness magnitude (b.), where the depths were calculated from the planetary radius $R_{P}$ and stellar radius $R_{\star}$. Note that $(R_{P}/R_{\star})^2$ is an easy to compute proxy for the transit depth.

Finally, we note that \hatcur\ is at an ecliptic latitude of $+3^{\circ}$, and is thus outside the field of view of the primary NASA {\em TESS} mission. It also was not observed during the {\em K2} mission. The discovery of this planet by HATNet demonstrates that in the era of wide-field space-based transit surveys, interesting planets amenable to detailed characterization remain to be discovered, even from the ground.


\acknowledgements 

\paragraph{Acknowledgements}
HATNet operations have been funded by NASA grants NNG04GN74G as well as NNX13AJ15G. Follow-up of HATNet targets has been partially supported through NSF grant AST-1108686.
B.L.\ is supported by the NSF Graduate Research Fellowship, grant no. DGE 1762114.
J.H.\ acknowledges support from NASA grant NNX14AE87G.
G.B., J.H.\ and W.B.\ acknowledge partial support from NASA grant NNX17AB61G.
K.P.\ acknowledges support from NASA grant 80NSSC18K1009.
I.B.\ thanks EuropeanCommunity’s Seventh Framework Programme (FP7/2007-2013) under grant agreement number RG226604 (OPTICON). and the Programme National dePlanétologie” (PNP) of CNRS/INSU
We acknowledge partial support from the {\em Kepler} Mission under NASA Cooperative Agreement NCC2-1390 (D.W.L., PI).
Data presented in this paper are based on observations obtained at the HAT station at the Submillimeter Array of SAO, and the HAT station at the Fred Lawrence Whipple Observatory of SAO.
We acknowledge the use of the AAVSO Photometric All-Sky Survey (APASS),
funded by the Robert Martin Ayers Sciences Fund, and the SIMBAD
database, operated at CDS, Strasbourg, France.
Data presented herein were obtained at the WIYN Observatory from telescope time allocated to NN-EXPLORE through the scientific partnership of the National Aeronautics and Space Administration, the National Science Foundation, and the National Optical Astronomy Observatory. This work was supported by a NASA WIYN PI Data Award, administered by the NASA Exoplanet Science Institute.
This work has made use of data from the European Space Agency (ESA)
mission {\it Gaia}\footnote{https://www.cosmos.esa.int/gaia}, processed by
the {\it Gaia} Data Processing and Analysis Consortium (DPAC,
\footnote{https://www.cosmos.esa.int/web/gaia/dpac/consortium}). Funding
for the DPAC has been provided by national institutions, in particular
the institutions participating in the {\it Gaia} Multilateral Agreement.
This research has made use of the NASA Exoplanet Archive\footnote{https://exoplanetarchive.ipac.caltech.edu/}, 
which is operated by the California Institute of Technology, 
under contract with the National Aeronautics and Space Administration 
under the Exoplanet Exploration Program.
The authors wish to recognize and acknowledge the very significant
cultural role and reverence that the summit of Mauna Kea has always
had within the indigenous Hawaiian community. We are most fortunate to
have the opportunity to conduct observations from this mountain.

\facilities{HATNet, FLWO:1.5m (TRES), ARC (ARCES), OHP:1.93m (Sophie),
  Keck:I (HIRES), FLWO:1.2m (KeplerCam), WIYN (DSSI), Gaia, Exoplanet
  Archive}

\software{FITSH \citep{pal:2012}, BLS \citep{kovacs:2002:BLS},
  VARTOOLS \citep{hartman:2016:vartools}, SPC
  \citep{buchhave:2012:spc}, MWDUST \citep{bovy:2016}, Astropy
  \citep{astropy:2013,astropy:2018}}

\bibliographystyle{aasjournal}
\bibliography{hatsbib}

\begin{thebibliography}{}
\expandafter\ifx\csname natexlab\endcsname\relax\def\natexlab#1{#1}\fi
\providecommand{\url}[1]{\href{#1}{#1}}

\bibitem[{{Astropy Collaboration} {et~al.}(2013){Astropy Collaboration},
  {Robitaille}, {Tollerud}, {Greenfield}, {Droettboom}, {Bray}, {Aldcroft},
  {Davis}, {Ginsburg}, {Price-Whelan}, {Kerzendorf}, {Conley}, {Crighton},
  {Barbary}, {Muna}, {Ferguson}, {Grollier}, {Parikh}, {Nair}, {Unther},
  {Deil}, {Woillez}, {Conseil}, {Kramer}, {Turner}, {Singer}, {Fox}, {Weaver},
  {Zabalza}, {Edwards}, {Azalee Bostroem}, {Burke}, {Casey}, {Crawford},
  {Dencheva}, {Ely}, {Jenness}, {Labrie}, {Lim}, {Pierfederici}, {Pontzen},
  {Ptak}, {Refsdal}, {Servillat}, \& {Streicher}}]{astropy:2013}
{Astropy Collaboration}, {Robitaille}, T.~P., {Tollerud}, E.~J., {et~al.} 2013,
  \aap, 558, A33

\bibitem[{{Auvergne} {et~al.}(2009){Auvergne}, {Bodin}, {Boisnard}, {Buey},
  {Chaintreuil}, {Epstein}, {Jouret}, {Lam-Trong}, {Levacher}, {Magnan},
  {Perez}, {Plasson}, {Plesseria}, {Peter}, {Steller}, {Tiph{\`e}ne}, {Baglin},
  {Agogu{\'e}}, {Appourchaux}, {Barbet}, {Beaufort}, {Bellenger}, {Berlin},
  {Bernardi}, {Blouin}, {Boumier}, {Bonneau}, {Briet}, {Butler}, {Cautain},
  {Chiavassa}, {Costes}, {Cuvilho}, {Cunha-Parro}, {de Oliveira Fialho},
  {Decaudin}, {Defise}, {Djalal}, {Docclo}, {Drummond}, {Dupuis}, {Exil},
  {Faur{\'e}}, {Gaboriaud}, {Gamet}, {Gavalda}, {Grolleau}, {Gueguen},
  {Guivarc'h}, {Guterman}, {Hasiba}, {Huntzinger}, {Hustaix}, {Imbert},
  {Jeanville}, {Johlander}, {Jorda}, {Journoud}, {Karioty}, {Kerjean},
  {Lafond}, {Lapeyrere}, {Landiech}, {Larqu{\'e}}, {Laudet}, {Le Merrer},
  {Leporati}, {Leruyet}, {Levieuge}, {Llebaria}, {Martin}, {Mazy}, {Mesnager},
  {Michel}, {Moalic}, {Monjoin}, {Naudet}, {Neukirchner}, {Nguyen-Kim},
  {Ollivier}, {Orcesi}, {Ottacher}, {Oulali}, {Parisot}, {Perruchot},
  {Piacentino}, {Pinheiro da Silva}, {Platzer}, {Pontet}, {Pradines},
  {Quentin}, {Rohbeck}, {Rolland}, {Rollenhagen}, {Romagnan}, {Russ}, {Samadi},
  {Schmidt}, {Schwartz}, {Sebbag}, {Smit}, {Sunter}, {Tello}, {Toulouse},
  {Ulmer}, {Vandermarcq}, {Vergnault}, {Wallner}, {Waultier}, \&
  {Zanatta}}]{auvergne:2009}
{Auvergne}, M., {Bodin}, P., {Boisnard}, L., {et~al.} 2009, \aap, 506, 411

\bibitem[{{Bakos} {et~al.}(2004){Bakos}, {Noyes}, {Kov{\'a}cs}, {Stanek},
  {Sasselov}, \& {Domsa}}]{bakos:2004:hatnet}
{Bakos}, G., {Noyes}, R.~W., {Kov{\'a}cs}, G., {et~al.} 2004, \pasp, 116, 266

\bibitem[{{Bakos}(2018)}]{bakos:2018:book}
{Bakos}, G.~{\'A}. 2018, {The HATNet and HATSouth Exoplanet Surveys}, 111

\bibitem[{{Bakos} {et~al.}(2010){Bakos}, {Torres}, {P{\'a}l}, {Hartman},
  {Kov{\'a}cs}, {Noyes}, {Latham}, {Sasselov}, {Sip{\H o}cz}, {Esquerdo},
  {Fischer}, {Johnson}, {Marcy}, {Butler}, {Isaacson}, {Howard}, {Vogt},
  {Kov{\'a}cs}, {Fernandez}, {Mo{\'o}r}, {Stefanik}, {L{\'a}z{\'a}r}, {Papp},
  \& {S{\'a}ri}}]{bakos:2010:hat11}
{Bakos}, G.~{\'A}., {Torres}, G., {P{\'a}l}, A., {et~al.} 2010, \apj, 710, 1724

\bibitem[{{Bakos} {et~al.}(2013){Bakos}, {Csubry}, {Penev}, {Bayliss},
  {Jord{\'a}n}, {Afonso}, {Hartman}, {Henning}, {Kov{\'a}cs}, {Noyes},
  {B{\'e}ky}, {Suc}, {Cs{\'a}k}, {Rabus}, {L{\'a}z{\'a}r}, {Papp}, {S{\'a}ri},
  {Conroy}, {Zhou}, {Sackett}, {Schmidt}, {Mancini}, {Sasselov}, \&
  {Ueltzhoeffer}}]{bakos:2013:hatsouth}
{Bakos}, G.~{\'A}., {Csubry}, Z., {Penev}, K., {et~al.} 2013, \pasp, 125, 154

\bibitem[{{Bakos} {et~al.}(2018){Bakos}, {Bayliss}, {Bento}, {Bhatti}, {Brahm},
  {Csubry}, {Espinoza}, {Hartman}, {Henning}, \&
  {Jord{\'a}n}}]{bakos:2018:hats71}
{Bakos}, G.~{\'A}., {Bayliss}, D., {Bento}, J., {et~al.} 2018, arXiv e-prints,
  arXiv:1812.09406

\bibitem[{{Bieryla} {et~al.}(2014){Bieryla}, {Hartman}, {Bakos}, {Bhatti},
  {Kov{\'a}cs}, {Boisse}, {Latham}, {Buchhave}, {Csubry}, {Penev}, {de
  Val-Borro}, {B{\'e}ky}, {Falco}, {Torres}, {Noyes}, {Berlind}, {Calkins},
  {Esquerdo}, {L{\'a}z{\'a}r}, {Papp}, \& {S{\'a}ri}}]{bieryla:2014:hat49}
{Bieryla}, A., {Hartman}, J.~D., {Bakos}, G.~{\'A}., {et~al.} 2014, \aj, 147,
  84

\bibitem[{{Boisse} {et~al.}(2013){Boisse}, {Hartman}, {Bakos}, {Penev},
  {Csubry}, {B{\'e}ky}, {Latham}, {Bieryla}, {Torres}, {Kov{\'a}cs},
  {Buchhave}, {Hansen}, {Everett}, {Esquerdo}, {Szklen{\'a}r}, {Falco},
  {Shporer}, {Fulton}, {Noyes}, {Stefanik}, {L{\'a}z{\'a}r}, {Papp}, \&
  {S{\'a}ri}}]{boisse:2013:hat4243}
{Boisse}, I., {Hartman}, J.~D., {Bakos}, G.~{\'A}., {et~al.} 2013, \aap, 558,
  A86

\bibitem[{{Borucki} {et~al.}(2010){Borucki}, {Koch}, {Basri}, {Batalha},
  {Brown}, {Caldwell}, {Caldwell}, {Christensen-Dalsgaard}, {Cochran},
  {DeVore}, {Dunham}, {Dupree}, {Gautier}, {Geary}, {Gilliland}, {Gould},
  {Howell}, {Jenkins}, {Kondo}, {Latham}, {Marcy}, {Meibom}, {Kjeldsen},
  {Lissauer}, {Monet}, {Morrison}, {Sasselov}, {Tarter}, {Boss}, {Brownlee},
  {Owen}, {Buzasi}, {Charbonneau}, {Doyle}, {Fortney}, {Ford}, {Holman},
  {Seager}, {Steffen}, {Welsh}, {Rowe}, {Anderson}, {Buchhave}, {Ciardi},
  {Walkowicz}, {Sherry}, {Horch}, {Isaacson}, {Everett}, {Fischer}, {Torres},
  {Johnson}, {Endl}, {MacQueen}, {Bryson}, {Dotson}, {Haas}, {Kolodziejczak},
  {Van Cleve}, {Chandrasekaran}, {Twicken}, {Quintana}, {Clarke}, {Allen},
  {Li}, {Wu}, {Tenenbaum}, {Verner}, {Bruhweiler}, {Barnes}, \&
  {Prsa}}]{borucki:2010}
{Borucki}, W.~J., {Koch}, D., {Basri}, G., {et~al.} 2010, Science, 327, 977

\bibitem[{{Bouchy} {et~al.}(2009){Bouchy}, {H{\'e}brard}, {Udry}, {Delfosse},
  {Boisse}, {Desort}, {Bonfils}, {Eggenberger}, {Ehrenreich}, {Forveille},
  {Lagrange}, {Le Coroller}, {Lovis}, {Moutou}, {Pepe}, {Perrier}, {Pont},
  {Queloz}, {Santos}, {S{\'e}gransan}, \& {Vidal-Madjar}}]{bouchy:2009}
{Bouchy}, F., {H{\'e}brard}, G., {Udry}, S., {et~al.} 2009, \aap, 505, 853

\bibitem[{{Bovy} {et~al.}(2016){Bovy}, {Rix}, {Green}, {Schlafly}, \&
  {Finkbeiner}}]{bovy:2016}
{Bovy}, J., {Rix}, H.-W., {Green}, G.~M., {Schlafly}, E.~F., \& {Finkbeiner},
  D.~P. 2016, \apj, 818, 130

\bibitem[{{Boyajian} {et~al.}(2012){Boyajian}, {von Braun}, {van Belle},
  {McAlister}, {ten Brummelaar}, {Kane}, {Muirhead}, {Jones}, {White},
  {Schaefer}, {Ciardi}, {Henry}, {L{\'o}pez-Morales}, {Ridgway}, {Gies}, {Jao},
  {Rojas-Ayala}, {Parks}, {Sturmann}, {Sturmann}, {Turner}, {Farrington},
  {Goldfinger}, \& {Berger}}]{boyajian:2012}
{Boyajian}, T.~S., {von Braun}, K., {van Belle}, G., {et~al.} 2012, \apj, 757,
  112

\bibitem[{Bressan {et~al.}(2012)Bressan, Marigo, Girardi, Salasnich, Dal~Cero,
  Rubele, \& Nanni}]{bressan:2012:parsec}
Bressan, A., Marigo, P., Girardi, L., {et~al.} 2012, \mnras, 427, 127

\bibitem[{{Buchhave} {et~al.}(2012){Buchhave}, {Latham}, {Johansen},
  {Bizzarro}, {Torres}, {Rowe}, {Batalha}, {Borucki}, {Brugamyer}, {Caldwell},
  {Bryson}, {Ciardi}, {Cochran}, {Endl}, {Esquerdo}, {Ford}, {Geary},
  {Gilliland}, {Hansen}, {Isaacson}, {Laird}, {Lucas}, {Marcy}, {Morse},
  {Robertson}, {Shporer}, {Stefanik}, {Still}, \& {Quinn}}]{buchhave:2012:spc}
{Buchhave}, L.~A., {Latham}, D.~W., {Johansen}, A., {et~al.} 2012, \nat, 486,
  375

\bibitem[{{Butler} {et~al.}(1996){Butler}, {Marcy}, {Williams}, {McCarthy},
  {Dosanjh}, \& {Vogt}}]{butler:1996}
{Butler}, R.~P., {Marcy}, G.~W., {Williams}, E., {et~al.} 1996, \pasp, 108, 500

\bibitem[{{Cardelli} {et~al.}(1989){Cardelli}, {Clayton}, \&
  {Mathis}}]{cardelli:1989}
{Cardelli}, J.~A., {Clayton}, G.~C., \& {Mathis}, J.~S. 1989, \apj, 345, 245

\bibitem[{{Charbonneau} {et~al.}(2002){Charbonneau}, {Brown}, {Noyes}, \&
  {Gilliland}}]{char:2002:atm}
{Charbonneau}, D., {Brown}, T.~M., {Noyes}, R.~W., \& {Gilliland}, R.~L. 2002,
  \apj, 568, 377

\bibitem[{{Chen} {et~al.}(2014{\natexlab{a}}){Chen}, {van Boekel}, {Wang},
  {Nikolov}, {Fortney}, {Seemann}, {Wang}, {Mancini}, \&
  {Henning}}]{chen:2014:wasp43}
{Chen}, G., {van Boekel}, R., {Wang}, H., {et~al.} 2014{\natexlab{a}}, \aap,
  563, A40

\bibitem[{{Chen} {et~al.}(2014{\natexlab{b}}){Chen}, {Girardi}, {Bressan},
  {Marigo}, {Barbieri}, \& {Kong}}]{chen:2014:parsec}
{Chen}, Y., {Girardi}, L., {Bressan}, A., {et~al.} 2014{\natexlab{b}}, \mnras,
  444, 2525

\bibitem[{{Claret}(2004)}]{claret:2004}
{Claret}, A. 2004, \aap, 428, 1001

\bibitem[{{Claret}(2018)}]{claret:2018}
---. 2018, \aap, 618, A20

\bibitem[{{Claret} {et~al.}(2012){Claret}, {Hauschildt}, \&
  {Witte}}]{claret:2012}
{Claret}, A., {Hauschildt}, P.~H., \& {Witte}, S. 2012, \aap, 546, A14

\bibitem[{{Claret} {et~al.}(2013){Claret}, {Hauschildt}, \&
  {Witte}}]{claret:2013}
---. 2013, \aap, 552, A16

\bibitem[{{Dawson} \& {Johnson}(2018)}]{dawson:2018}
{Dawson}, R.~I., \& {Johnson}, J.~A. 2018, \araa, 56, 175

\bibitem[{{F\H{u}resz}(2008)}]{furesz:2008}
{F\H{u}resz}, G. 2008, PhD thesis, {Univ. of Szeged, Hungary}

\bibitem[{{Gaia Collaboration} {et~al.}(2016){Gaia Collaboration}, {Brown},
  {Vallenari}, {Prusti}, {de Bruijne}, {Mignard}, {Drimmel}, {Babusiaux},
  {Bailer-Jones}, {Bastian}, \& et~al.}]{gaiadr1}
{Gaia Collaboration}, {Brown}, A.~G.~A., {Vallenari}, A., {et~al.} 2016, \aap,
  595, A2

\bibitem[{{Gaia Collaboration} {et~al.}(2018){Gaia Collaboration}, {Brown},
  {Vallenari}, {Prusti}, {de Bruijne}, {Babusiaux}, {Bailer-Jones}, {Biermann},
  {Evans}, {Eyer}, {Jansen}, {Jordi}, {Klioner}, {Lammers}, {Lindegren},
  {Luri}, {Mignard}, {Panem}, {Pourbaix}, {Randich}, {Sartoretti}, {Siddiqui},
  {Soubiran}, {van Leeuwen}, {Walton}, {Arenou}, {Bastian}, {Cropper},
  {Drimmel}, {Katz}, {Lattanzi}, {Bakker}, {Cacciari}, {Casta{\~n}eda},
  {Chaoul}, {Cheek}, {De Angeli}, {Fabricius}, {Guerra}, {Holl}, {Masana},
  {Messineo}, {Mowlavi}, {Nienartowicz}, {Panuzzo}, {Portell}, {Riello},
  {Seabroke}, {Tanga}, {Th{\'e}venin}, {Gracia-Abril}, {Comoretto},
  {Garcia-Reinaldos}, {Teyssier}, {Altmann}, {Andrae}, {Audard},
  {Bellas-Velidis}, {Benson}, {Berthier}, {Blomme}, {Burgess}, {Busso},
  {Carry}, {Cellino}, {Clementini}, {Clotet}, {Creevey}, {Davidson}, {De
  Ridder}, {Delchambre}, {Dell'Oro}, {Ducourant},
  {Fern{\'a}ndez-Hern{\'a}ndez}, {Fouesneau}, {Fr{\'e}mat}, {Galluccio},
  {Garc{\'\i}a-Torres}, {Gonz{\'a}lez-N{\'u}{\~n}ez}, {Gonz{\'a}lez-Vidal},
  {Gosset}, {Guy}, {Halbwachs}, {Hambly}, {Harrison}, {Hern{\'a}ndez},
  {Hestroffer}, {Hodgkin}, {Hutton}, {Jasniewicz}, {Jean-Antoine-Piccolo},
  {Jordan}, {Korn}, {Krone-Martins}, {Lanzafame}, {Lebzelter}, {L{\"o}ffler},
  {Manteiga}, {Marrese}, {Mart{\'\i}n-Fleitas}, {Moitinho}, {Mora}, {Muinonen},
  {Osinde}, {Pancino}, {Pauwels}, {Petit}, {Recio-Blanco}, {Richards},
  {Rimoldini}, {Robin}, {Sarro}, {Siopis}, {Smith}, {Sozzetti}, {S{\"u}veges},
  {Torra}, {van Reeven}, {Abbas}, {Abreu Aramburu}, {Accart}, {Aerts},
  {Altavilla}, {{\'A}lvarez}, {Alvarez}, {Alves}, {Anderson}, {Andrei},
  {Anglada Varela}, {Antiche}, {Antoja}, {Arcay}, {Astraatmadja}, {Bach},
  {Baker}, {Balaguer-N{\'u}{\~n}ez}, {Balm}, {Barache}, {Barata}, {Barbato},
  {Barblan}, {Barklem}, {Barrado}, {Barros}, {Barstow}, {Bartholom{\'e}
  Mu{\~n}oz}, {Bassilana}, {Becciani}, {Bellazzini}, {Berihuete}, {Bertone},
  {Bianchi}, {Bienaym{\'e}}, {Blanco-Cuaresma}, {Boch}, {Boeche}, {Bombrun},
  {Borrachero}, {Bossini}, {Bouquillon}, {Bourda}, {Bragaglia}, {Bramante},
  {Breddels}, {Bressan}, {Brouillet}, {Br{\"u}semeister}, {Brugaletta},
  {Bucciarelli}, {Burlacu}, {Busonero}, {Butkevich}, {Buzzi}, {Caffau},
  {Cancelliere}, {Cannizzaro}, {Cantat-Gaudin}, {Carballo}, {Carlucci},
  {Carrasco}, {Casamiquela}, {Castellani}, {Castro-Ginard}, {Charlot},
  {Chemin}, {Chiavassa}, {Cocozza}, {Costigan}, {Cowell}, {Crifo}, {Crosta},
  {Crowley}, {Cuypers}, {Dafonte}, {Damerdji}, {Dapergolas}, {David}, {David},
  {de Laverny}, {De Luise}, {De March}, {de Martino}, {de Souza}, {de Torres},
  {Debosscher}, {del Pozo}, {Delbo}, {Delgado}, {Delgado}, {Di Matteo},
  {Diakite}, {Diener}, {Distefano}, {Dolding}, {Drazinos}, {Dur{\'a}n},
  {Edvardsson}, {Enke}, {Eriksson}, {Esquej}, {Eynard Bontemps}, {Fabre},
  {Fabrizio}, {Faigler}, {Falc{\~a}o}, {Farr{\`a}s Casas}, {Federici},
  {Fedorets}, {Fernique}, {Figueras}, {Filippi}, {Findeisen}, {Fonti},
  {Fraile}, {Fraser}, {Fr{\'e}zouls}, {Gai}, {Galleti}, {Garabato},
  {Garc{\'\i}a-Sedano}, {Garofalo}, {Garralda}, {Gavel}, {Gavras}, {Gerssen},
  {Geyer}, {Giacobbe}, {Gilmore}, {Girona}, {Giuffrida}, {Glass}, {Gomes},
  {Granvik}, {Gueguen}, {Guerrier}, {Guiraud}, {Guti{\'e}rrez-S{\'a}nchez},
  {Haigron}, {Hatzidimitriou}, {Hauser}, {Haywood}, {Heiter}, {Helmi}, {Heu},
  {Hilger}, {Hobbs}, {Hofmann}, {Holland}, {Huckle}, {Hypki}, {Icardi},
  {Jan{\ss}en}, {Jevardat de Fombelle}, {Jonker}, {Juh{\'a}sz}, {Julbe},
  {Karampelas}, {Kewley}, {Klar}, {Kochoska}, {Kohley}, {Kolenberg},
  {Kontizas}, {Kontizas}, {Koposov}, {Kordopatis}, {Kostrzewa-Rutkowska},
  {Koubsky}, {Lambert}, {Lanza}, {Lasne}, {Lavigne}, {Le Fustec}, {Le
  Poncin-Lafitte}, {Lebreton}, {Leccia}, {Leclerc}, {Lecoeur-Taibi},
  {Lenhardt}, {Leroux}, {Liao}, {Licata}, {Lindstr{\o}m}, {Lister}, {Livanou},
  {Lobel}, {L{\'o}pez}, {Managau}, {Mann}, {Mantelet}, {Marchal}, {Marchant},
  {Marconi}, {Marinoni}, {Marschalk{\'o}}, {Marshall}, {Martino}, {Marton},
  {Mary}, {Massari}, {Matijevi{\v{c}}}, {Mazeh}, {McMillan}, {Messina},
  {Michalik}, {Millar}, {Molina}, {Molinaro}, {Moln{\'a}r}, {Montegriffo},
  {Mor}, {Morbidelli}, {Morel}, {Morris}, {Mulone}, {Muraveva}, {Musella},
  {Nelemans}, {Nicastro}, {Noval}, {O'Mullane}, {Ord{\'e}novic},
  {Ord{\'o}{\~n}ez-Blanco}, {Osborne}, {Pagani}, {Pagano}, {Pailler},
  {Palacin}, {Palaversa}, {Panahi}, {Pawlak}, {Piersimoni}, {Pineau}, {Plachy},
  {Plum}, {Poggio}, {Poujoulet}, {Pr{\v{s}}a}, {Pulone}, {Racero}, {Ragaini},
  {Rambaux}, {Ramos-Lerate}, {Regibo}, {Reyl{\'e}}, {Riclet}, {Ripepi}, {Riva},
  {Rivard}, {Rixon}, {Roegiers}, {Roelens}, {Romero-G{\'o}mez}, {Rowell},
  {Royer}, {Ruiz-Dern}, {Sadowski}, {Sagrist{\`a} Sell{\'e}s}, {Sahlmann},
  {Salgado}, {Salguero}, {Sanna}, {Santana-Ros}, {Sarasso}, {Savietto},
  {Schultheis}, {Sciacca}, {Segol}, {Segovia}, {S{\'e}gransan}, {Shih},
  {Siltala}, {Silva}, {Smart}, {Smith}, {Solano}, {Solitro}, {Sordo}, {Soria
  Nieto}, {Souchay}, {Spagna}, {Spoto}, {Stampa}, {Steele},
  {Steidelm{\"u}ller}, {Stephenson}, {Stoev}, {Suess}, {Surdej}, {Szabados},
  {Szegedi-Elek}, {Tapiador}, {Taris}, {Tauran}, {Taylor}, {Teixeira},
  {Terrett}, {Teyssand ier}, {Thuillot}, {Titarenko}, {Torra Clotet}, {Turon},
  {Ulla}, {Utrilla}, {Uzzi}, {Vaillant}, {Valentini}, {Valette}, {van Elteren},
  {Van Hemelryck}, {van Leeuwen}, {Vaschetto}, {Vecchiato}, {Veljanoski},
  {Viala}, {Vicente}, {Vogt}, {von Essen}, {Voss}, {Votruba}, {Voutsinas},
  {Walmsley}, {Weiler}, {Wertz}, {Wevers}, {Wyrzykowski}, {Yoldas},
  {{\v{Z}}erjal}, {Ziaeepour}, {Zorec}, {Zschocke}, {Zucker}, {Zurbach}, \&
  {Zwitter}}]{gaiadr2}
---. 2018, \aap, 616, A1

\bibitem[{{Gaudi} {et~al.}(2005){Gaudi}, {Seager}, \&
  {Mallen-Ornelas}}]{gaudi:2005}
{Gaudi}, B.~S., {Seager}, S., \& {Mallen-Ornelas}, G. 2005, \apj, 623, 472

\bibitem[{{Hansen} \& {Barman}(2007)}]{hansen:2007}
{Hansen}, B.~M.~S., \& {Barman}, T. 2007, \apj, 671, 861

\bibitem[{{Hartman} \& {Bakos}(2016)}]{hartman:2016:vartools}
{Hartman}, J.~D., \& {Bakos}, G.~{\'A}. 2016, Astronomy and Computing, 17, 1

\bibitem[{{Hartman} {et~al.}(2011){Hartman}, {Bakos}, {Noyes}, {Sip{\H o}cz},
  {Kov{\'a}cs}, {Mazeh}, {Shporer}, \& {P{\'a}l}}]{hartman:2011:kmdwarf}
{Hartman}, J.~D., {Bakos}, G.~{\'A}., {Noyes}, R.~W., {et~al.} 2011, \aj, 141,
  166

\bibitem[{{Hartman} {et~al.}(2012){Hartman}, {Bakos}, {B{\'e}ky}, {Torres},
  {Latham}, {Csubry}, {Penev}, {Shporer}, {Fulton}, {Buchhave}, {Johnson},
  {Howard}, {Marcy}, {Fischer}, {Kov{\'a}cs}, {Noyes}, {Esquerdo}, {Everett},
  {Szklen{\'a}r}, {Quinn}, {Bieryla}, {Knox}, {Hinz}, {Sasselov}, {F{\H
  u}r{\'e}sz}, {Stefanik}, {L{\'a}z{\'a}r}, {Papp}, \&
  {S{\'a}ri}}]{hartman:2012:hat39hat41}
{Hartman}, J.~D., {Bakos}, G.~{\'A}., {B{\'e}ky}, B., {et~al.} 2012, \aj, 144,
  139

\bibitem[{{Hartman} {et~al.}(2015){Hartman}, {Bhatti}, {Bakos}, {Bieryla},
  {Kov{\'a}cs}, {Latham}, {Csubry}, {de Val-Borro}, {Penev}, {Buchhave},
  {Torres}, {Howard}, {Marcy}, {Johnson}, {Isaacson}, {Sato}, {Boisse},
  {Falco}, {Everett}, {Szklenar}, {Fulton}, {Shporer}, {Kov{\'a}cs}, {Hansen},
  {B{\'e}ky}, {Noyes}, {L{\'a}z{\'a}r}, {Papp}, \&
  {S{\'a}ri}}]{hartman:2015:hat50hat53}
{Hartman}, J.~D., {Bhatti}, W., {Bakos}, G.~{\'A}., {et~al.} 2015, \aj, 150,
  168

\bibitem[{Hartman {et~al.}(2016)Hartman, Bakos, Bhatti, Penev, Bieryla, Latham,
  Kov{\'a}cs, Torres, Csubry, Val-Borro, \& et~al.}]{hartman:2016:hat65hat66}
Hartman, J.~D., Bakos, G.~{\'A}., Bhatti, W., {et~al.} 2016, \aj, 152, 182

\bibitem[{{Hartman} {et~al.}(2019){Hartman}, {Bakos}, {Bayliss}, {Bento},
  {Bhatti}, {Brahm}, {Csubry}, {Espinoza}, {Henning}, {Jord{\'a}n}, {Mancini},
  {Penev}, {Rabus}, {Sarkis}, {Suc}, {de Val-Borro}, {Zhou}, {Addison},
  {Arriagada}, {Butler}, {Crane}, {Durkan}, {Shectman}, {Tan}, {Thompson},
  {Tinney}, {Wright}, {L{\'a}z{\'a}r}, {Papp}, \&
  {S{\'a}ri}}]{hartman:2019:hats69}
{Hartman}, J.~D., {Bakos}, G.~{\'A}., {Bayliss}, D., {et~al.} 2019, \aj, 157,
  55

\bibitem[{{Henden} {et~al.}(2009){Henden}, {Welch}, {Terrell}, \&
  {Levine}}]{henden:2009}
{Henden}, A.~A., {Welch}, D.~L., {Terrell}, D., \& {Levine}, S.~E. 2009, in
  American Astronomical Society Meeting Abstracts, Vol. 214, American
  Astronomical Society Meeting Abstracts \#214, \#407.02

\bibitem[{{Horch} {et~al.}(2011){Horch}, {van Altena}, {Howell}, {Sherry}, \&
  {Ciardi}}]{horch:2011}
{Horch}, E.~P., {van Altena}, W.~F., {Howell}, S.~B., {Sherry}, W.~H., \&
  {Ciardi}, D.~R. 2011, \aj, 141, 180

\bibitem[{{Horch} {et~al.}(2009){Horch}, {Veillette}, {Baena Gall{\'e}},
  {Shah}, {O'Rielly}, \& {van Altena}}]{horch:2009}
{Horch}, E.~P., {Veillette}, D.~R., {Baena Gall{\'e}}, R., {et~al.} 2009, \aj,
  137, 5057

\bibitem[{{Howard} {et~al.}(2010){Howard}, {Johnson}, {Marcy}, {Fischer},
  {Wright}, {Bernat}, {Henry}, {Peek}, {Isaacson}, {Apps}, {Endl}, {Cochran},
  {Valenti}, {Anderson}, \& {Piskunov}}]{howard:2010:cps}
{Howard}, A.~W., {Johnson}, J.~A., {Marcy}, G.~W., {et~al.} 2010, \apj, 721,
  1467

\bibitem[{{Howell} {et~al.}(2011){Howell}, {Everett}, {Sherry}, {Horch}, \&
  {Ciardi}}]{howell:2011}
{Howell}, S.~B., {Everett}, M.~E., {Sherry}, W., {Horch}, E., \& {Ciardi},
  D.~R. 2011, \aj, 142, 19

\bibitem[{{Howell} {et~al.}(2014){Howell}, {Sobeck}, {Haas}, {Still},
  {Barclay}, {Mullally}, {Troeltzsch}, {Aigrain}, {Bryson}, {Caldwell},
  {Chaplin}, {Cochran}, {Huber}, {Marcy}, {Miglio}, {Najita}, {Smith},
  {Twicken}, \& {Fortney}}]{howell:2014}
{Howell}, S.~B., {Sobeck}, C., {Haas}, M., {et~al.} 2014, \pasp, 126, 398

\bibitem[{{Kirk} {et~al.}(2016){Kirk}, {Wheatley}, {Louden}, {Littlefair},
  {Copperwheat}, {Armstrong}, {Marsh}, \& {Dhillon}}]{kirk:2016:wasp52}
{Kirk}, J., {Wheatley}, P.~J., {Louden}, T., {et~al.} 2016, \mnras, 463, 2922

\bibitem[{Kirk {et~al.}(2017)Kirk, Wheatley, Louden, Skillen, King, McCormac,
  \& Irwin}]{kirk:2017:wasp80}
Kirk, J., Wheatley, P.~J., Louden, T., {et~al.} 2017, \mnras, 474, 876

\bibitem[{{Kov{\'a}cs} {et~al.}(2005){Kov{\'a}cs}, {Bakos}, \&
  {Noyes}}]{kovacs:2005:TFA}
{Kov{\'a}cs}, G., {Bakos}, G., \& {Noyes}, R.~W. 2005, \mnras, 356, 557

\bibitem[{{Kov{\'a}cs} {et~al.}(2002){Kov{\'a}cs}, {Zucker}, \&
  {Mazeh}}]{kovacs:2002:BLS}
{Kov{\'a}cs}, G., {Zucker}, S., \& {Mazeh}, T. 2002, \aap, 391, 369

\bibitem[{{Louden} {et~al.}(2017){Louden}, {Wheatley}, {Irwin}, {Kirk}, \&
  {Skillen}}]{louden:2017:wasp52}
{Louden}, T., {Wheatley}, P.~J., {Irwin}, P. G.~J., {Kirk}, J., \& {Skillen},
  I. 2017, \mnras, 470, 742

\bibitem[{{Mancini} {et~al.}(2014){Mancini}, {Southworth}, {Ciceri}, {Dominik},
  {Henning}, {J{\o}rgensen}, {Lanza}, {Rabus}, {Snodgrass}, {Vilela},
  {Alsubai}, {Bozza}, {Bramich}, {Calchi Novati}, {D'Ago}, {Figuera Jaimes},
  {Galianni}, {Gu}, {Harps{\o}e}, {Hinse}, {Hundertmark}, {Juncher}, {Kains},
  {Korhonen}, {Popovas}, {Rahvar}, {Skottfelt}, {Street}, {Surdej}, {Tsapras},
  {Wang}, \& {Wertz}}]{mancini:2014:wasp80}
{Mancini}, L., {Southworth}, J., {Ciceri}, S., {et~al.} 2014, \aap, 562, A126

\bibitem[{{Mandel} \& {Agol}(2002)}]{mandel:2002}
{Mandel}, K., \& {Agol}, E. 2002, \apjl, 580, L171

\bibitem[{{Mayor} \& {Queloz}(1995)}]{mayor:1995:exo}
{Mayor}, M., \& {Queloz}, D. 1995, \nat, 378, 355

\bibitem[{Morton \& Winn(2014)}]{morton:2014}
Morton, T.~D., \& Winn, J.~N. 2014, \apj, 796, 47

\bibitem[{{Nielsen} {et~al.}(2020){Nielsen}, {Brahm}, {Bouchy}, {Espinoza},
  {Turner}, {Rappaport}, {Pearce}, {Ricker}, {Vanderspek}, {Latham}, {Seager},
  {Winn}, {Jenkins}, {Acton}, {Bakos}, {Barclay}, {Barkaoui}, {Bhatti},
  {Brice{\~n}o}, {Bryant}, {Burleigh}, {Ciardi}, {Collins}, {Collins}, {Cooke},
  {Csubry}, {dos Santos}, {Eigm{\"u}ller}, {Fausnaugh}, {Gan}, {Gillon},
  {Goad}, {Guerrero}, {Hagelberg}, {Hart}, {Henning}, {Huang}, {Jehin},
  {Jenkins}, {Jord{\'a}n}, {Kielkopf}, {Kossakowski}, {Lavie}, {Law}, {Lendl},
  {de Leon}, {Lovis}, {Mann}, {Marmier}, {McCormac}, {Mori}, {Moyano},
  {Narita}, {Osip}, {Otegi}, {Pepe}, {Pozuelos}, {Raynard}, {Relles}, {Sarkis},
  {S{\'e}gransan}, {Seidel}, {Shporer}, {Stalport}, {Stockdale}, {Suc},
  {Tamura}, {Tan}, {Tilbrook}, {Ting}, {Trifonov}, {Udry}, {Vanderburg},
  {Wheatley}, {Wingham}, {Zhan}, \& {Ziegler}}]{nielsen:2020}
{Nielsen}, L.~D., {Brahm}, R., {Bouchy}, F., {et~al.} 2020, \aap, 639, A76

\bibitem[{{P{\'a}l}(2012)}]{pal:2012}
{P{\'a}l}, A. 2012, \mnras, 421, 1825

\bibitem[{Pollacco {et~al.}(2006)Pollacco, Skillen, Cameron, Christian,
  Hellier, Irwin, Lister, Street, West, Anderson, Clarkson, Deeg, Enoch, Evans,
  Fitzsimmons, Haswell, Hodgkin, Horne, Kane, Keenan, Maxted, Norton, Osborne,
  N.R.Parley, Ryans, Smalley, Wheatley, \& Wilson}]{pollacco:2006}
Pollacco, D., Skillen, I., Cameron, A.~C., {et~al.} 2006, \pasp, 118, 1407

\bibitem[{{Price-Whelan} {et~al.}(2018){Price-Whelan}, {Sip{\H{o}}cz},
  {G{\"u}nther}, {Lim}, {Crawford}, {Conseil}, {Shupe}, {Craig}, {Dencheva},
  {Ginsburg}, {VanderPlas}, {Bradley}, {P{\'e}rez-Su{\'a}rez}, {de Val-Borro},
  {Paper Contributors}, {Aldcroft}, {Cruz}, {Robitaille}, {Tollerud},
  {Coordination Committee}, {Ardelean}, {Babej}, {Bach}, {Bachetti}, {Bakanov},
  {Bamford}, {Barentsen}, {Barmby}, {Baumbach}, {Berry}, {Biscani}, {Boquien},
  {Bostroem}, {Bouma}, {Brammer}, {Bray}, {Breytenbach}, {Buddelmeijer},
  {Burke}, {Calderone}, {Cano Rodr{\'\i}guez}, {Cara}, {Cardoso}, {Cheedella},
  {Copin}, {Corrales}, {Crichton}, {D{\textquoteright}Avella}, {Deil},
  {Depagne}, {Dietrich}, {Donath}, {Droettboom}, {Earl}, {Erben}, {Fabbro},
  {Ferreira}, {Finethy}, {Fox}, {Garrison}, {Gibbons}, {Goldstein}, {Gommers},
  {Greco}, {Greenfield}, {Groener}, {Grollier}, {Hagen}, {Hirst}, {Homeier},
  {Horton}, {Hosseinzadeh}, {Hu}, {Hunkeler}, {Ivezi{\'c}}, {Jain}, {Jenness},
  {Kanarek}, {Kendrew}, {Kern}, {Kerzendorf}, {Khvalko}, {King}, {Kirkby},
  {Kulkarni}, {Kumar}, {Lee}, {Lenz}, {Littlefair}, {Ma}, {Macleod},
  {Mastropietro}, {McCully}, {Montagnac}, {Morris}, {Mueller}, {Mumford},
  {Muna}, {Murphy}, {Nelson}, {Nguyen}, {Ninan}, {N{\"o}the}, {Ogaz}, {Oh},
  {Parejko}, {Parley}, {Pascual}, {Patil}, {Patil}, {Plunkett}, {Prochaska},
  {Rastogi}, {Reddy Janga}, {Sabater}, {Sakurikar}, {Seifert}, {Sherbert},
  {Sherwood-Taylor}, {Shih}, {Sick}, {Silbiger}, {Singanamalla}, {Singer},
  {Sladen}, {Sooley}, {Sornarajah}, {Streicher}, {Teuben}, {Thomas},
  {Tremblay}, {Turner}, {Terr{\'o}n}, {van Kerkwijk}, {de la Vega}, {Watkins},
  {Weaver}, {Whitmore}, {Woillez}, {Zabalza}, \& {Contributors}}]{astropy:2018}
{Price-Whelan}, A.~M., {Sip{\H{o}}cz}, B.~M., {G{\"u}nther}, H.~M., {et~al.}
  2018, \aj, 156, 123

\bibitem[{{Randich} {et~al.}(2018){Randich}, {Tognelli}, {Jackson}, {Jeffries},
  {Degl'Innocenti}, {Pancino}, {Re Fiorentin}, {Spagna}, {Sacco}, {Bragaglia},
  {Magrini}, {Prada Moroni}, {Alfaro}, {Franciosini}, {Morbidelli},
  {Roccatagliata}, {Bouy}, {Bravi}, {Jim{\'e}nez-Esteban}, {Jordi}, {Zari},
  {Tautvai{\v{s}}iene}, {Drazdauskas}, {Mikolaitis}, {Gilmore}, {Feltzing},
  {Vallenari}, {Bensby}, {Koposov}, {Korn}, {Lanzafame}, {Smiljanic}, {Bayo},
  {Carraro}, {Costado}, {Heiter}, {Hourihane}, {Jofr{\'e}}, {Lewis}, {Monaco},
  {Prisinzano}, {Sbordone}, {Sousa}, {Worley}, \& {Zaggia}}]{randich:2018}
{Randich}, S., {Tognelli}, E., {Jackson}, R., {et~al.} 2018, \aap, 612, A99

\bibitem[{Ricker {et~al.}(2014)Ricker, Winn, Vanderspek, Latham, Bakos, Bean,
  Berta-Thompson, Brown, Buchhave, Butler, Butler, Chaplin, Charbonneau,
  Christensen-Dalsgaard, Clampin, Deming, Doty, Lee, Dressing, Dunham, Endl,
  Fressin, Ge, Henning, Holman, Howard, Ida, Jenkins, Jernigan, Johnson,
  Kaltenegger, Kawai, Kjeldsen, Laughlin, Levine, Lin, Lissauer, MacQueen,
  Marcy, McCullough, Morton, Narita, Paegert, Palle, Pepe, Pepper, Quirrenbach,
  Rinehart, Sasselov, Sato, Seager, Sozzetti, Stassun, Sullivan, Szentgyorgyi,
  Torres, Udry, \& Villasenor}]{ricker:2014}
Ricker, G.~R., Winn, J.~N., Vanderspek, R., {et~al.} 2014, Journal of
  Astronomical Telescopes, Instruments, and Systems, 1, 1

\bibitem[{{Sestovic} {et~al.}(2018){Sestovic}, {Demory}, \&
  {Queloz}}]{sestovic:2018}
{Sestovic}, M., {Demory}, B.-O., \& {Queloz}, D. 2018, \aap, 616, A76

\bibitem[{{Skrutskie} {et~al.}(2006){Skrutskie}, {Cutri}, {Stiening},
  {Weinberg}, {Schneider}, {Carpenter}, {Beichman}, {Capps}, {Chester},
  {Elias}, {Huchra}, {Liebert}, {Lonsdale}, {Monet}, {Price}, {Seitzer},
  {Jarrett}, {Kirkpatrick}, {Gizis}, {Howard}, {Evans}, {Fowler}, {Fullmer},
  {Hurt}, {Light}, {Kopan}, {Marsh}, {McCallon}, {Tam}, {Van Dyk}, \&
  {Wheelock}}]{skrutskie:2006}
{Skrutskie}, M.~F., {Cutri}, R.~M., {Stiening}, R., {et~al.} 2006, \aj, 131,
  1163

\bibitem[{Stassun {et~al.}(2017)Stassun, Collins, \& Gaudi}]{stassun:2017}
Stassun, K.~G., Collins, K.~A., \& Gaudi, B.~S. 2017, \aj, 153, 136

\bibitem[{{ter Braak}(2006)}]{terbraak:2006}
{ter Braak}, C.~J.~F. 2006, Statistics and Computing, 16, 239

\bibitem[{{Torres} {et~al.}(2007){Torres}, {Bakos}, {Kov{\'a}cs}, {Latham},
  {Fern{\'a}ndez}, {Noyes}, {Esquerdo}, {Sozzetti}, {Fischer}, {Butler},
  {Marcy}, {Stefanik}, {Sasselov}, {L{\'a}z{\'a}r}, {Papp}, \&
  {S{\'a}ri}}]{torres:2007:hat3}
{Torres}, G., {Bakos}, G.~{\'A}., {Kov{\'a}cs}, G., {et~al.} 2007, \apjl, 666,
  L121

\bibitem[{{Vogt} {et~al.}(1994){Vogt}, {Allen}, {Bigelow}, {Bresee}, {Brown},
  {Cantrall}, {Conrad}, {Couture}, {Delaney}, {Epps}, {Hilyard}, {Hilyard},
  {Horn}, {Jern}, {Kanto}, {Keane}, {Kibrick}, {Lewis}, {Osborne},
  {Pardeilhan}, {Pfister}, {Ricketts}, {Robinson}, {Stover}, {Tucker}, {Ward},
  \& {Wei}}]{vogt:1994}
{Vogt}, S.~S., {Allen}, S.~L., {Bigelow}, B.~C., {et~al.} 1994, in Society of
  Photo-Optical Instrumentation Engineers (SPIE) Conference Series, Vol. 2198,
  Society of Photo-Optical Instrumentation Engineers (SPIE) Conference Series,
  ed. D.~L. {Crawford} \& E.~R. {Craine}, 362

\bibitem[{{Wang} {et~al.}(2003){Wang}, {Hildebrand}, {Hobbs}, {Heimsath},
  {Kelderhouse}, {Loewenstein}, {Lucero}, {Rockosi}, {Sandford}, {Sundwall},
  {Thorburn}, \& {York}}]{wang:2003}
{Wang}, S.-i., {Hildebrand}, R.~H., {Hobbs}, L.~M., {et~al.} 2003, in Society
  of Photo-Optical Instrumentation Engineers (SPIE) Conference Series, Vol.
  4841, Instrument Design and Performance for Optical/Infrared Ground-based
  Telescopes, ed. M.~{Iye} \& A.~F.~M. {Moorwood}, 1145--1156

\bibitem[{{Weaver} {et~al.}(2020){Weaver}, {L{\'o}pez-Morales}, {Espinoza},
  {Rackham}, {Osip}, {Apai}, {Jord{\'a}n}, {Bixel}, {Lewis}, {Alam}, {Kirk},
  {McGruder}, {Rodler}, \& {Fienco}}]{weaver:2020:wasp43}
{Weaver}, I.~C., {L{\'o}pez-Morales}, M., {Espinoza}, N., {et~al.} 2020, \aj,
  159, 13

\bibitem[{{Wright} {et~al.}(2010){Wright}, {Eisenhardt}, {Mainzer}, {Ressler},
  {Cutri}, {Jarrett}, {Kirkpatrick}, {Padgett}, {McMillan}, {Skrutskie},
  {Stanford}, {Cohen}, {Walker}, {Mather}, {Leisawitz}, {Gautier}, {McLean},
  {Benford}, {Lonsdale}, {Blain}, {Mendez}, {Irace}, {Duval}, {Liu}, {Royer},
  {Heinrichsen}, {Howard}, {Shannon}, {Kendall}, {Walsh}, {Larsen}, {Cardon},
  {Schick}, {Schwalm}, {Abid}, {Fabinsky}, {Naes}, \& {Tsai}}]{wright:2010}
{Wright}, E.~L., {Eisenhardt}, P. R.~M., {Mainzer}, A.~K., {et~al.} 2010, \aj,
  140, 1868

\bibitem[{{Yi} {et~al.}(2001){Yi}, {Demarque}, {Kim}, {Lee}, {Ree}, {Lejeune},
  \& {Barnes}}]{yi:2001}
{Yi}, S., {Demarque}, P., {Kim}, Y.-C., {et~al.} 2001, \apjs, 136, 417

\bibitem[{{Zacharias} {et~al.}(2013){Zacharias}, {Finch}, {Girard}, {Henden},
  {Bartlett}, {Monet}, \& {Zacharias}}]{zacharias:2013:ucac4}
{Zacharias}, N., {Finch}, C.~T., {Girard}, T.~M., {et~al.} 2013, \aj, 145, 44

\bibitem[{{Zechmeister} \& {K{\"u}rster}(2009)}]{zechmeister:2009}
{Zechmeister}, M., \& {K{\"u}rster}, M. 2009, \aap, 496, 577

\end{thebibliography}

\end{document}